\documentclass[twocolumn,english,pra,showpacs,aps,superscriptaddress]{revtex4-1}

\usepackage[T1]{fontenc}
\usepackage[latin9]{inputenc}
\usepackage{amsmath}
\usepackage{amssymb}
\usepackage{graphicx}
\usepackage{times}
\usepackage{helvet}
\usepackage{mathrsfs}
\usepackage{babel}

\usepackage[hidelinks,colorlinks]{hyperref}
\hypersetup{citecolor=[rgb]{0.25,0.14,0.63}}
\hypersetup{urlcolor=[rgb]{0.25,0.14,0.63}}
\hypersetup{linkcolor=black}

\makeatletter
 
 \@ifundefined{textcolor}{}
 {%
   \definecolor{BLACK}{gray}{0}
   \definecolor{WHITE}{gray}{1}
   \definecolor{RED}{rgb}{1,0,0}
   \definecolor{GREEN}{rgb}{0,1,0}
   \definecolor{BLUE}{rgb}{0,0,1}
   \definecolor{CYAN}{cmyk}{1,0,0,0}
   \definecolor{MAGENTA}{cmyk}{0,1,0,0}
   \definecolor{YELLOW}{cmyk}{0,0,1,0}
 }

\@ifundefined{textcolor}{}{%
 \definecolor{BLACK}{gray}{0}
 \definecolor{WHITE}{gray}{1}
 \definecolor{RED}{rgb}{1,0,0}
 \definecolor{GREEN}{rgb}{0,1,0}
 \definecolor{BLUE}{rgb}{0,0,1}
 \definecolor{CYAN}{cmyk}{1,0,0,0}
 \definecolor{MAGENTA}{cmyk}{0,1,0,0}
 \definecolor{YELLOW}{cmyk}{0,0,1,0}
 }

\@ifundefined{definecolor}{\@ifundefined{definecolor}
 {\@ifundefined{definecolor}
 {\@ifundefined{definecolor}
 {\@ifundefined{definecolor}
 {\@ifundefined{definecolor}
 {\usepackage{color}}{}
}{}
}{}
}{}
}{}
}{}
\makeatother

\begin{document}

\title{Atomic twin-beams and violation of a motional-state Bell inequality from a phase-fluctuating quasi-condensate source}

\author{R.~J.~Lewis-Swan}
\affiliation{JILA, NIST and University of Colorado, 440 UCB, Boulder, Colorado 80309, USA}
\affiliation{Center for Theory of Quantum Matter, University of Colorado, Boulder, Colorado 80309, USA}
\author{K.~V.~Kheruntsyan}
\affiliation{The University of Queensland, School of Mathematics and Physics,
Brisbane, Queensland 4072, Australia}

\date{\today }
\begin{abstract}
We investigate the dynamics of atomic twin beams produced from a phase-fluctuating source, specifically a 1D Bose gas in the quasi-condensate regime, motivated by the experiment reported in \textit{Nature Physics} {\bf 7}, 608 (2011). A short-time analytic model is constructed, which is a modified version of the undepleted pump approximation widely used in quantum and atom optics, except that here we take into account the initial phase fluctuations of the pump source as opposed to assuming long-range phase coherence.
We use this model to make quantitative and qualitative predictions of how phase-fluctuations of the source impact the two-particle correlations of scattered atom-pairs. The model is benchmarked against detailed numerical simulations using stochastic phase-space methods, and is shown to validate the intuitive notion that the broadening of momentum-space correlation functions between atoms scattered from a quasi-condensate is driven by the broadened momentum width of the source compared to a true phase coherent condensate. Finally, we combine these theoretical tools and results to investigate the effect phase fluctuations of the twin-beam source can have on a proposed demonstration of a violation of a Bell inequality, which intrinsically relies on phase-sensitive pair correlations. 
\end{abstract}


\maketitle

\section{Introduction}
The creation, manipulation and application of correlated twin-atoms is a topic of interest across a range of cold atoms experiments. This is driven by their potential utility in quantum technologies such as precision atom interferometry \cite{Klempt_TwinBeams_2011,Bucker_TwinBeams_2011,Linnemann_SU11_2016,Klempt_Clock2016,Cronin_2009} and quantum simulation \cite{Clark_2017,Feng_2017,Hu_2019}, as well as fundamental tests of quantum mechanics such as atomic EPR entanglement \cite{Peise_EPR_2015,OberthalerEPR_2018,KlemptEPR_2018}, the atomic Hong-Ou-Mandel effect \cite{Perrier_2019,Lopes_HOM_2015,RLS_HOM_2014} and demonstrations of a Bell inequality using motional degrees of freedom and massive particles \cite{KK_SubPoiss_2010,KK_MultimodeCS_2012,Kofler_EPR_2012,Dussarrat_Interferometer_2017,RLS_Bell_2015}. Essential to each of these applications has been the ability to well characterise, both theoretically and experimentally, the nature of these atom-pairs as well as their intrinsic correlations \cite{Perrin_AtomicFWM_2008,Perrin_AtomPairs_2007,Hodgman_Microscope_2017,Bonneau_DoubleWell_2017,Wasak_BogRegions_2014,Trippenbach_Perturbative_2008,Bucker_ParametricMatterWaves_2012,Perrier_2019,Ogren_Correlations_2009,Deuar_NumberSqueeze_2013,Zin_2019,Deuar_2007,Deuar_2014,Deuar_2011,Deuar_2011_Bog} 

Recent experiments involving atom-pairs have relied on protocols which can be reduced to the archetypal process of four-wave mixing: A pair of atoms in a coherent Bose-Einstein condensate (BEC) interact and are 
scattered into a distinct pair of modes (in terms of either spatial, motional or internal degrees of freedom) outside the condensate. Consequently, the vast majority of theoretical work in the literature pertaining to twin-atom production has focused on this case of scattering from a coherent source \cite{Ogren_Correlations_2009,Perrin_AtomicFWM_2008,RLS_HOM_2014,RLS_Bell_2015,Trippenbach_Perturbative_2008,Wasak_BogRegions_2014}. In particular, the coherent source is often replaced by a classical field to make the resulting model more tractable. However, a series of recent experiments have evolved this paradigm by utilizing Bose gases in the quasi-condensate regime as a source for the scattered pairs \cite{Bucker_TwinBeams_2011,KK_MultimodeCS_2012,Bonneau_CorrelatedBeams_2013}. A quasi-condensate lacks the long-range order and phase-coherence of a BEC, and so is not immediately ammenable to the same approximations, such as replacement by a classical field. This presents a motivation to construct more detailed and sophisticated models of pair-production which can specifically account for the phase fluctuations of a quasi-condensate source \cite{Bucker_ParametricMatterWaves_2012,Wasak_RamanScattering_2012,Bucker_TwinBeams_2011,Zin_2019}.

Here, we construct a simple analytic theory which describes the short-time limit of pair-production from a phase-fluctuating 1D Bose gas source. The choice of a 1D quasi-condensate source is motivated specifically by the experiment of Ref.~\cite{Bucker_TwinBeams_2011} wherein correlated atomic twin beams were produced using a parametrically excited $1$D quasi-condensate. Our theoretical analysis enables quantitative and qualitative insight into second-order (density-density) correlations between the twin beams, specifically the scaling of the peak correlation strength and correlation widths as a function of the temperature of the source quasi-condensate. As a consequence of this analysis, we quantitatively establish the connection between the broadening of the correlation width and the width of the momentum distribution of the source quasi-condensate, a result which was previously experimentally observed in Ref.~\cite{KK_MultimodeCS_2012} for the related example of two quasi-condensates in the $3$D collisional regime. As a benchmark, we compare the predicted results to numerical simulations using stochastic phase-space methods and find excellent agreement with respect to the scaling of correlations with the temperature of the source quasi-condensate. Furthermore, we demonstrate that some qualitative predictions of the analytic model remain a useful guide beyond the short-time limit of the theory. 

Our ultimate interest in the properties of twin beams produced via phase-fluctuating sources is motivated by their possible utility for fundamental tests of quantum mechanics such as violation of a Bell inequality \cite{RLS_Bell_2015,Dussarrat_Interferometer_2017,Bonneau_DoubleWell_2017}. Demonstration of a Bell inequality violation intrinsically relies on phase-sensitive measurements of pair correlations with respect to well-defined relative phase-settings of the underlying interferometric setup \cite{Dussarrat_Interferometer_2017,Rarity_Bell_1990}. For ultracold atoms, this was shown to be theoretically possible using a phase-coherent BEC source \cite{RLS_Bell_2015}. The question of whether a similar violation is possible with a phase-fluctuating quasi-condensate source is far from trivial, and naive arguments might suggest an a priori negative answer given that the relative phase-fluctuations of the source may degrade or completely destroy the necessary phase-sensitive correlations between the twin beams. However, our detailed calculations instead indicate that whereas the phase-fluctuations of the quasi-condensate do 
indeed degrade the amplitude of the phase-sensitive pair correlations (relative to a coherent source), they remain sufficiently strong to enable Bell inequality violation for a sufficiently cold quasi-condensate source. 

The derivations of the analytic models and numerical calculations in this manuscript are tailored towards the specific experimental setup of Ref.~\cite{Bucker_TwinBeams_2011}, however, our results can be generalized to 
related models of atom-pair production. In particular, our analytic model and proposed protocol to demonstrate a Bell inequality violation could be readily adapted to describe $s$-wave scattering from elongated 3D quasi-condensates \cite{KK_MultimodeCS_2012,Zin_2019,Petrov_2001}.

The paper is structured as follows. First, we outline the effective model of the pair production process from a quasi-condensate source in Sec.~\ref{sec:PairProduction}. Next, in Sec.~\ref{sec:ShortTimeTheory} 
we use a perturbative treatment to derive approximate analytic expressions for the momentum-space pair correlation functions of the twin-beams. In Sec.~\ref{sec:NumericalSims} we then validate these analytic expressions by comparing to 
detailed numerical calculations based on the positive-$P$ stochastic phase-space method, whilst also investigating beyond the short-time validity of the analytic treatment. Finally, we build on the preceding results and 
investigate the feasibility of demonstrating a violation of a Bell inequality with a quasi-condensate source in Sec.~\ref{sec:BellTest}. We conclude the paper with summarizing remarks in Sec.~\ref{sec:Conclusion}.

\section{Model \label{sec:PairProduction}}

\subsection{Pair production process}
Our theoretical model of the pair production process begins from the generic Hamiltonian describing a dilute 3D degenerate Bose gas trapped in a potential $V(\mathbf{r})$ with $s$-wave contact interactions,
\begin{eqnarray}
 \hat{H} & = & \int ~ d\mathbf{r} ~ \Big\{ \hat{\psi}^{\dagger}(\mathbf{r}) \left[ \frac{-\hbar^2}{2m}\nabla^2 + V(\mathbf{r}) \right] \hat{\psi}(\mathbf{r}) \notag \\
 & &   + \frac{g}{2}\hat{\psi}^{\dagger}(\mathbf{r})\hat{\psi}^{\dagger}(\mathbf{r})\hat{\psi}(\mathbf{r})\hat{\psi}(\mathbf{r}) \Big\} . \label{eqn:BosonHam}
\end{eqnarray}
Here, $g = 4\pi\hbar^2 a_s/m$ characterises the strength of interactions with $s$-wave scattering length $a_s$ and atomic mass $m$. 

The Hamiltonian Eq.~(\ref{eqn:BosonHam}) can be simplified by the physical considerations of the experimental setup of Ref.~\cite{Bucker_TwinBeams_2011}. In particular, the Bose gas is trapped in an elongated cylindrically symmetric harmonic potential with sufficiently large and equal trapping frequencies in the transverse directions (here taken to be along $y$ and $z$, with $\omega_y=\omega_z\equiv\omega_{\perp}$) so that we may derive an effective one-dimensional (1D) model of the system involving only explicit spatial dependence along the weakly confined direction $x$. The experimental sequence of Ref.~\cite{Bucker_TwinBeams_2011} entails parametrically driving the trapping potential along $y$ according to optimal control theory \cite{Bucker_OptimalControl_2013} such that the quasi-condensate formed initially in the transverse ground state $(n_y, n_z)=(0,0)$ is coherently transferred to the first excited state $(n_y, n_z)=(1,0)$ of the transverse trapping potential. Here, $n_{y,z}$ denotes the energy levels of the transverse harmonic potential. The dynamics of the preparation protocol was previously investigated in detail in Ref.~\cite{Bucker_OptimalControl_2013}. For simplicity, we ignore this transfer stage in our model and assume that the quasi-condensate, with the same equilibrium configuration along $x$ as before, is simply formed in the $(n_y, n_z)=(1,0)$ transverse state.

We continue by expanding the field operator as $\hat{\psi}(\mathbf{r}) \equiv \sum_{m_x,n_y,n_z} \hat{a}_{m_x,n_y,n_z} \varphi_{m_x}(x) \phi^{(ho)}_{n_y}(y) \phi^{(ho)}_{n_z}(z)$. Here, $\phi^{(ho)}_{n_y}(y)$ [$\phi^{(ho)}_{n_z}(z)$] is the harmonic oscillator basis function of the $n_y$th ($n_z$th) mode of the trapping potential, $\varphi_{m_x}(x)$ are a set of basis functions along the $x$ dimension labeled by independent indices $m_x$, and $\hat{a}_{m_x,n_y,n_z}$ is the corresponding bosonic creation operator for the mode.

As there are no dynamics along the $z$ dimension we restrict the expansion to $n_z = 0$ throughout the remainder of the manuscript and suppress the associated subscript. Moreover, we assume that only the levels $n_y=0,1$ are involved in the state preparation and de-excitation process, motivated by the experimental inclusion of a small anharmonicity in the trapping potential along $y$. Such anharmonicity ensures that excitation to higher energy levels is energetically suppressed, so that the levels with $n_y \geq 2$ are never significantly populated and can be ignored in our model. Substitution of the expansion of the field operator back into Eq.~(\ref{eqn:BosonHam}) under these conditions allows us to integrate out the $y$ and $z$ dimensions to yield an effective $1$D Hamiltonian. Specifically, defining the new field operators $\hat{\psi}_i(x) = \sum_{m_x} \hat{a}_{m_x, i, 0} \varphi_{m_x}(x)$, with $i\equiv n_y=\{0,1\}$, the $1$D Hamiltonian can be expressed as $\hat{H} = \hat{H}_0 + \hat{H}_{\mathrm{int}}$, where
\begin{eqnarray}
 \hat{H}_0 & = & \int dx ~ \sum_{i=0,1} \hat{\psi}^{\dagger}_i(x) \bigg[ \frac{-\hbar^2}{2m}\frac{\partial^2}{\partial x^2} \notag \\ 
 & &  + V(x) + \delta_{i,1}\hbar\omega_y \bigg] \hat{\psi}_i(x)
\label{eqn:H0}
\end{eqnarray}
is the single-body Hamiltonian, and 
\begin{eqnarray}
 \hat{H}_{\mathrm{int}} & = & \!\int~dx \bigg\{ \frac{g_{00}}{2} \hat{\psi}^{\dagger}_{0}(x) \hat{\psi}^{\dagger}_{0}(x) \hat{\psi}_{0}(x) \hat{\psi}_0(x) \notag \\
 &  + & \frac{g_{11}}{2} \hat{\psi}^{\dagger}_{1}(x) \hat{\psi}^{\dagger}_{1}(x) \hat{\psi}_{1}(x) \hat{\psi}_{1}(x) \notag \\
 &  + & 2g_{01} \hat{\psi}^{\dagger}_{0}(x) \hat{\psi}_{0}(x) \,\hat{\psi}^{\dagger}_{1}(x) \hat{\psi}_{1}(x)  \notag \\
 &  + & \frac{g_{01}}{2} \big[ \hat{\psi}^{\dagger}_{1}(x) \hat{\psi}^{\dagger}_{1}(x) \hat{\psi}_{0}(x) \hat{\psi}_{0}(x)  + h. c. \big] \bigg\} 
\label{eqn:Hamiltonian}
\end{eqnarray}
is the interaction Hamiltonian, where $h.c.$ refers to the Hermitian conjugate. As the oscillator basis states can be taken to be real-valued functions, the effective 1D coupling strengths are determined by $g_{ij} = \int dz [\phi^{(ho)}_0(z)]^4 \int dy [\phi^{(ho)}_i(y)]^2 [\phi^{(ho)}_j(y)]^2$. Thus we have $g_{00} = g/(2\pi L_{y}L_{z})$, $g_{01} = g/(4\pi L_{y}L_{z})$ and $g_{11} = 3g/(8\pi L_{y}L_{z})$ where $L_{y(z)} = \sqrt{\hbar/(m\omega_{y(z)})}$ is the harmonic oscillator length in the $y$ ($z$) direction.

We can identify the last line of the interaction Hamiltonian $\hat{H}_{\mathrm{int}}$, Eq.~(\ref{eqn:Hamiltonian}), as describing the well-known process of bosonic pair-production via four-wave mixing:
\begin{equation}
 \hat{H}_{\mathrm{FWM}} = \frac{g_{01}}{2} \int dx ~ \hat{\psi}^{\dagger}_{1}(x) \hat{\psi}^{\dagger}_{1}(x) \hat{\psi}_{0}(x) \hat{\psi}_{0}(x) + h.c. . \label{eqn:H_4WM}
\end{equation}
This effective Hamiltonian $\hat{H}_{\mathrm{FWM}}$ encapsulate the physics of the collisional de-excitation process, by which two atoms in the transverse excited state 
($n_y=1$) relax to the transverse ground state ($n_y=0$). In this conversion process, the excess potential energy $2\hbar\omega_y$ from the strongly confining trap along $y$  is converted into kinetic energy along the weakly trapped longitudinal direction $x$ [see Eq.~(\ref{eqn:H0})]. 
The remaining three terms of Eq.~(\ref{eqn:Hamiltonian}) describe elastic scattering and can be interpreted as spatially and time dependent mean-field potentials which the scattered atoms move in \cite{Deuar_NumberSqueeze_2013}. As the energetic contributions of the elastic scattering and the weak trapping potential along $x$ are small relative to the excess potential energy $2\hbar\omega_y$, we can assume that these terms remain approximately constant and hence amount to a constant phase shift which can be ignored. 
Under this assumption, and due to conservation of energy and approximately the momentum,
the atom pairs scattered from the initial transverse excited state $n_y=1$ (where the atoms are at rest longitudinally) into $n_y=0$ state will then have counter-propagating longitudinal momenta $k_x \approx \pm k_0$ where $k_0 = \sqrt{2m\omega_y/\hbar}$.

\subsection{Correlation functions}

In the simplest case of a true condensate source in the undepleted pump  approximation, the Hamiltonian Eq.~(\ref{eqn:H_4WM}) describes the widely studied phenomena of spontaneous optical parametric down-conversion. This process is known to produce the two-mode squeezed vacuum state \cite{walls2008quantum,lewis2016ultracold}, which exibits strong correlations between the down-converted particles. Whilst this simplification is not valid in the system under investigation, due to the absence of a true condensate in 1D, we still expect strong, non-trivial correlations in the down-converted field due to the pair-wise nature of the scattering process. 

The correlations between the atoms scattered into the transverse ground state $n_y = 0$ according to Eq.~(\ref{eqn:H_4WM}) can be characterized via Glauber's normalized second-order correlation function \cite{Glauber_1963},
\begin{equation}
    g^{(2)}(k,k',t) = \frac{\langle \hat{a}^{\dagger}(k,t) \hat{a}^{\dagger}(k',t) \hat{a}(k') \hat{a}(k,t) \rangle}{\langle \hat{a}^{\dagger}(k,t)\hat{a}(k,t) \rangle \langle \hat{a}^{\dagger}(k',t)\hat{a}(k',t) \rangle} , \label{eqn:g2_defn}
\end{equation}
which describes normally-ordered density-density correlations between momentum modes $k$ and $k'$. Here, $\hat{a}(k,t)$ [$\hat{a}^{\dagger}(k,t)$] is the Fourier transform component of the down-converted field $\hat{\psi}_0(x,t)$ [$\hat{\psi}_0^{\dagger}(x,t)$] at time $t$, corresponding to the annihilation (creation) operator 
for an atom in momentum mode $k$. The normalization of Eq.~(\ref{eqn:g2_defn}) is introduced to imply that the result $g^{(2)}(k,k,t) = 1$ corresponds to the absence of any correlation between the modes $k$ and $k'$. Due to energy and approximate momentum conservation in the scattering 
process, we expect a non-trivial `back-to-back' (BB) correlation, $g^{(2)}(k,k',t)> 1$  for $k' \approx -k$ \cite{Perrin_AtomicFWM_2008}. Similarly, we also characterize the auto-correlation of the beams by the `collinear' 
(CL) correlation for $k' \approx k$.

\subsection{Quasi-condensate source}
\label{Quasi-condensate source} 

In Ref.~\cite{Bucker_TwinBeams_2011} it was reported that the initial cloud (before excited-state transfer) was characterised as a 1D quasi-condensate of temperature $T\lesssim 40$~nK. To model this initial condition for atoms in the excited state as well (owing to the coherent nature of the transfer protocol), we employ a Luttinger liquid approach wherein $\hat{\psi}_1(x,0) \equiv \hat{\psi}_1(x)= \sqrt{\rho(x)}\mathrm{exp}[i\hat{\phi}(x)]$ \cite{Bouchoule_1DBoseGas_2012,Cazalilla_Bosonization_2004}. Here, we ignore density fluctuations and so $\rho(x) = \rho_0(1-x^2/R^2)$ is the initial density profile of the quasi-condensate in the Thomas-Fermi (TF) approximation with TF radius $R$ and peak density $\rho_0$, 
and $\hat{\phi}(x)$ characterises the phase-profile of the quasi-condensate \cite{Petrov_1DqBEC_2000}.

In contrast to a true BEC, an equilibrium 1D quasi-condensate at temperature $T$ is characterized by a lack of long-range order, with the one-body density matrix given by 
$G^{(1)}(x,x') \equiv \langle \hat{\psi}^{\dagger}_1(x)\hat{\psi}_1(x') \rangle = \sqrt{\rho(x)\rho(x')} \mathrm{exp}[-\langle (\delta \hat{\phi}_{xx'}) \rangle /2]$ where 
$\delta\hat{\phi}_{xx'} \equiv \hat{\phi}(x)-\hat{\phi}(x')$. At experimentally relevant temperature scales, the phase fluctuations are dominated by the thermal contribution of low-energy phonon excitations \cite{Petrov_1DqBEC_2000} and are given by
\begin{equation}
 \langle \delta \hat{\phi}_{xx'} \rangle \simeq \frac{\mathcal{A}}{2} \mathrm{log} \left| \frac{(1-x'/R_x)(1+x/R_x)}{(1+x'/R_x)(1-x/R_x)} \right| , \label{eqn:PhaseFluct_T}
\end{equation}
where $\mathcal{A} = (2g_{11}k_B T)/[R(\hbar\omega_x)^2]$.
In the limits of $x,x' \ll R_x$ or $x-x' \ll R_x$ this expression can be approximated by the simpler form $\langle \delta \hat{\phi}_{xx'} \rangle \simeq |x-x'|/l_T$, where the thermal phase coherence length $l_T \equiv R_x/\mathcal{A} = \hbar^2\rho_0/(m k_B T)$ is equivalent to the result for a uniform 1D quasi-condensate at density $\rho_0$.

A second important feature of the quasi-condensate---and which is important to the pair-production process discussed in this article---is the broadening of the momentum distribution 
with temperature. The momentum distribution for a trapped quasi-condensate can be evaluated as \cite{Kheruntsyan_FiniteT1D_2005}
\begin{equation}
 n_1(k) \simeq \frac{1}{\pi}\int dx \frac{2l_T(x)\rho(x)}{1 + (2l_T(x)k)^2},
\end{equation}
within the local density approximation (LDA), where $l_T(x) \equiv \hbar^2\rho(x)/(m k_B T)$ is the local phase coherence length. The LDA assumes that each infinitesimal slice of the trapped gas behaves locally like a uniform gas of fixed density $ \rho(x)$, 
and the momentum distribution of the total trapped gas is then given by the summation of the uniform result for each slice of the gas, characterised by the Lorentzian integrand.  
Whilst this integral cannot be solved exactly when $\rho(x)$ is given by the Thomas-Fermi approximation, by assuming that the dominant contribution will be from the center of the cloud 
with $\rho(x) \simeq \rho_0$, then we may use the result of a uniform quasi-condensate to estimate the half-width at half-maximum (HWHM) $w_k$ of $n(k)$ to scale with temperature as $w_k \sim 1/l_T \propto T$.

\section{Short-time analytic treatment of pair production 
\label{sec:ShortTimeTheory}}

To gain a better understanding of how the phase-fluctuations of the quasi-condensate source affects the pair-production process and impacts the resulting atom-atom correlations, we treat the system using a 
short-time analytic model previously outlined in Ref.~\cite{Ogren_Correlations_2009} in the context of spontaneous four-wave mixing via collisions of pure 3D condensates. We note that a 
key difference to the previous implementation of this technique is that it would be unjustified to make the usual mean-field replacement $\hat{\psi}_1(x) \rightarrow \langle \hat{\psi}_1(x) \rangle$ in the Hamiltonian of Eq.~(\ref{eqn:Hamiltonian}) due to the absence of a true long-range order in 1D. As such the intrinsically phase fluctuating nature of the source will prove to have important consequences.

First, we assume that, for sufficiently short times and sufficiently high population, depletion of the transversely excited quasi-condensate source (due to conversion to transverse ground state) may be ignored. This means that the only time-dependence of the field operator $\hat{\psi}_1(x,t)$ is the trivial accrual of a spatially independent phase, $\hat{\psi}_1(x,t)=\hat{\psi}_1(x)e^{i\omega_yt}$, due to the excitation energy in the transverse excited state (where we ignore the zero point energy). The stationary component $\hat{\psi}_1(x)$ is treated as in the Luttinger liquid approach, as specified earlier, wherein we ignore the quasi-condensate density fluctuations, while retaining its phase fluctuations, so that $\hat{\psi}_1(x,t) = \sqrt{\rho(x)}e^{i\hat{\phi}(x)-i\omega_y t}$. This is somewhat similar to the conventional undepleted pump approximation frequently used in quantum and atom optics, where the field operator describing a true condensate source is replaced by a static $c$-number corresponding to the mean-field. Here, however, we relax the mean-field approach and instead retain an operator description in order to correctly account for the phase-fluctuations. This assumption is crucial to our model and serves as an extension of the undepleted pump formalism beyond the coherent BEC paradigm.

Next, the transverse excitation energy $2\hbar\omega_y$ of the source (or the pump mode) atoms is assumed to be the dominant contribution to the energy of the scattered (down-converted) atom pairs. Accordingly, we can also ignore the energy shifts due to elastic $s$-wave interactions in $\hat{H}_{\mathrm{int}}$ [first three lines of Eq.~(\ref{eqn:Hamiltonian})] as small contributions that are not relevant to the dynamics of the down-converted field $\hat{\psi}_1(x)$. Finally, for developing the approximate analytic model of the current section, we ignore for simplicity the longitudinal trapping potential $V(x)$ as well.
(In the full numerical simulations of Sec.~\ref{sec:NumericalSims}, we keep this trapping potential term.) Collectively, these approximations mean that the dynamics of the field operator describing the scattered pairs, $\hat{\psi}_0(x,t)$, is governed by the kinetic energy term from the noninteracting Hamiltonian, Eq.~({\ref{eqn:H0}}), plus the effective four-wave mixing Hamiltonian $\hat{H}_{\mathrm{4WM}}$, Eq.~(\ref{eqn:H_4WM}), with the substitution of $\hat{\psi}_1(x,t) = \sqrt{\rho(x)}e^{i\hat{\phi}(x)-i\omega_y t}$.

We can now proceed to solve for the dynamics of the scattered atoms. Specifically, the Heisenberg equation of motion describing the field operator of the scattered atoms $\hat{\psi}_0(x,t)$ is given by
\begin{equation}
 \frac{\partial \hat{\psi}_0(x,t)}{\partial t} = \frac{i\hbar}{2m}\nabla^2\hat{\psi}_0(x,t) - i \hat{g}(x) e^{-2i\omega_yt} \hat{\psi}^{\dagger}_0(x,t) ,
 \label{Heisenberg1}
\end{equation} 
where we have introduced an effective phase-fluctuating coupling
\begin{equation}
    \hat{g}(x) \equiv \frac{g_{01}}{\hbar}[\hat{\psi}_1(x)]^2 = \frac{g_{01}}{\hbar}\rho(x)e^{2i\hat{\phi}(x)} ,
\end{equation} 
describing the inelastic scattering of atoms from 
the $n_y=1$ excited (pump) state to $n_y = 0$ ground (down-converted) state.

Moving to a rotating frame, $\hat{\psi}_0(x,t) \rightarrow \hat{\psi}_0(x,t)e^{i\omega_y t}$, we can cast this equation in the following form
\begin{equation}
 \frac{\partial \hat{\psi}_0(x,t)}{\partial t} = i\Big(\frac{\hbar }{2m}\nabla^2+\omega_y\Big)\hat{\psi}_0(x,t) - i \hat{g}(x)  \hat{\psi}^{\dagger}_0(x,t) ,
 \label{Heisenberg2}
\end{equation} 
which we note is equivalent [except for the phase-fluctuating nature of the effective coupling $\hat{g}(x)$] to the equation of motion describing the production of pair correlated atoms through a collision of two coherent BECs \cite{Ogren_Correlations_2009}, or through dissociation of a BEC of molecular dimers \cite{Ogren_dissociation_2008}, in the undepleted pump approximation. In those processes, the role of the effective detuning $\hbar\omega_y$ is taken by the collisional kinetic energy per atom $\hbar^2Q^2/2m$ (where $Q$ is the collision momentum), or half the energy mismatch $2\hbar|\Delta|$ between the free two-atom state in the dissociation threshold and the energy of the bound molecular state.

Transforming next to Fourier space, with
$\hat{\psi}_0(x,t) = \int dk ~ \hat{a}(k,t) e^{ikx} / \sqrt{2\pi}$, one can write the corresponding equation of motion for the Fourier component $\hat{a}(k,t)$ as
\begin{equation}
 \frac{d\hat{a}(k,t)}{dt} = -i\Delta_k\hat{a}(k,t) - i\int \frac{dq}{\sqrt{2\pi}} ~ \hat{\tilde{g}}(q+k) \hat{a}^{\dagger}(k,t) ,
\end{equation}
where $\Delta_k \!=\! \hbar k^2/(2m)\! -\! \omega_y$ and $\hat{\tilde{g}}(k) \!=\! \frac{1}{\sqrt{2\pi}}\int \!dx ~\hat{g}(x) e^{-ikx} $ is the Fourier transform of the effective coupling.

Following Ref. \cite{Ogren_Correlations_2009}, we write $\hat{a}(k,t)$ in terms of a Taylor expansion in $t$,
\begin{equation}
 \hat{a}(k,t) = \hat{a}(k,0) + \frac{\partial\hat{a}(k,t)}{\partial t} \bigg|_{t=0} t + \mathcal{O}(t^2) , \label{eqn:ShortTime_taylor}
\end{equation}
which is valid for $t \ll \hbar/[g_{01}\rho(0)]$ \cite{Ogren_Correlations_2009}. Evaluation of the relevant correlation functions then proceeds by substitution of this 
expansion into, \textit{e.g.}, Eq.~(\ref{eqn:g2_defn}), and use of the equal-time  bosonic commutation relations 
$[\hat{a}(k,t),\hat{a}^{\dagger}(k',t)] = \delta(k-k')$.

\subsection{Momentum-space density of the twin-atom beams}
Following this procedure the first-order correlation $n(k,k',t) \equiv \langle \hat{a}^{\dagger}(k,t) \hat{a}(k',t) \rangle$ is then, to lowest order in $t$,
\begin{eqnarray}
 n(k,k',t) & \simeq & t^2 \int \frac{dq}{2\pi} ~ \langle \hat{\tilde{g}}^{\dagger}(q+k) \hat{\tilde{g}}(q+k') \rangle , \notag \\
 & = & t^2 \int \frac{dx}{2\pi} e^{-i(k-k')x} \langle \hat{g}^{\dagger}(x) \hat{g}(x) \rangle . \label{eqn:n_ShortTime_int}
\end{eqnarray}
Here, $\langle \hat{g}^{\dagger}(x) \hat{g}(x) \rangle \equiv [g_{01}\rho(x)/\hbar]^2$ is insensitive to the phase-fluctuations of the quasi-condensate and is equivalent to the result for a true BEC in the undepleted pump approximation \cite{Ogren_Correlations_2009}. We are naturally interested in the case $k^{\prime} \approx k$, for which $n(k,k'=k,t)$ describes the momentum-space density of the scattered atoms $n(k,t)$. 
Thus, considering $k\approx k'$ we define $\Delta{k} = k'-k$ and evaluate the integral in Eq.~(\ref{eqn:n_ShortTime_int}), yielding
\begin{equation}
 n(k,k'=k+\Delta k,t) \simeq \left[ \frac{g_{01}\rho(0)t}{\hbar} \right]^2 \frac{R_x}{\sqrt{\pi}} \frac{J_{5/2}(\Delta{k}R_x)}{\left(\frac{\Delta{k}R_x}{2}\right)^{5/2}} , \label{eqn:n_kkdash_ShortTime}
\end{equation}
where $J_{\alpha}(x)$ is a Bessel function of the first kind of order $\alpha$. The momentum-space density profile of the scattered atoms is found by simplifying to the case $k=k'$, giving 
\begin{equation}
 n(k,t) = \frac{R_x}{\Gamma(7/2)\sqrt{\pi}} \left[\frac{g_{01}\rho(0)t}{\hbar}\right]^2  , \label{eqn:n_kk_ShortTime}
\end{equation}
which indicates that for short times the scattered atoms uniformly populate all possible momentum components.

\subsection{Atom-atom correlations in the short-time approximation}

In prior work, such as Ref.~\cite{Ogren_Correlations_2009}, the calculation of second-order correlation functions was simplified by the use of Wick's theorem, wherein the factorizability of second-order correlations means that they can be constructed as products of first-order correlations, $\langle \hat{a}^{\dagger}(k,t)\hat{a}(k',t)\rangle$ and $\langle \hat{a}(k,t)\hat{a}(k',t)\rangle$. However, Wick's theorem is only strictly valid in the case where the Hamiltonian driving the dynamics is no more than quadratic in creation and annihilation operators. While this is true in the conventional undepleted pump approximation, where the pump mode is replaced by a $c$-number \cite{Ogren_Correlations_2009}, in our case an operator description of the pump mode is retained to describe the phase-fluctuations and so the Hamiltonian Eq.~(\ref{eqn:H_4WM}) remains quartic and hence Wick's factorization scheme doesn't apply. This means that second-order correlations should be calculated by directly substituting the expansion Eq.~(\ref{eqn:ShortTime_taylor}) into the definition of Eq.~(\ref{eqn:g2_defn}). We emphasize this point as calculation of only first-order correlations and subsequent application of Wick's theorem in our case would lead to a misleading and incorrect result for the BB correlation.

We proceed to calculate the momentum space atom-atom correlations by evaluating the numerator of Eq.~(\ref{eqn:g2_defn}) with the expression Eq.~(\ref{eqn:ShortTime_taylor}) and retaining terms up to $\mathcal{O}(t^4)$ to obtain 
\begin{multline}
G^{(2)}(k,k',t)  \simeq  \frac{t^2}{2\pi} \langle \hat{\tilde{g}}^{\dagger}(k+k')\hat{\tilde{g}}(k+k') \rangle  \\
+ \frac{t^4}{(2\pi)^2} \iint dq dq' ~ \langle \hat{\tilde{g}}^{\dagger}(q+k)\hat{\tilde{g}}(q+k) \hat{\tilde{g}}^{\dagger}(q'+k')\hat{\tilde{g}}(q'+k')\rangle \\
+ \frac{t^4}{(2\pi)^2}  \iint dq dq' ~ \langle \hat{\tilde{g}}^{\dagger}(q+k)\hat{\tilde{g}}(q+k')  \hat{\tilde{g}}^{\dagger}(q'+k')\hat{\tilde{g}}(q'+k) \rangle,  \label{eqn:ShortTime_G2}
\end{multline}
where $G^{(2)}(k,k',t)=\langle \hat{a}^{\dagger}(k,t) \hat{a}^{\dagger}(k',t) \hat{a}(k') \hat{a}(k,t) \rangle$ is the unnormalized second-order correlation. Note that here we have assumed that at $t=0$ the pump and scattered modes are uncorrelated, \textit{i.e.}, $\langle \hat{\tilde{g}}(k)\hat{a}(k^{\prime},0) \rangle = \langle \hat{\tilde{g}}(k)\rangle \langle \hat{a}(k^{\prime},0) \rangle$.

By direct substitution of $\hat{\tilde{g}}(x)=\sqrt{\rho(x)}e^{i\hbar\hat{\phi}(x)}$ the expression for $G^{(2)}(k,k',t)$ may be simplified to two non-trivial cases, specifically relating to the previously defined BB ($k'\approx -k$) and CL ($k' \approx k$) correlations that we are most interested in. In particular, Eq.~(\ref{eqn:ShortTime_G2}) can be rewritten as
\begin{multline}
 G^{(2)}(k,k',t) \simeq n(k,t)n(k',t) + |n(k,k',t)|^2 \\
 + \Big(\frac{g_{01}t}{2\pi\hbar}\Big)^2  \!\!\!\iint \!dx dx'  e^{-i(k+k')(x-x')} \!\rho(x)\rho(x') e^{-2\langle (\delta\hat{\phi}_{xx'})^2 \rangle} . \label{eqn:ShortTime_G2_step2}
\end{multline}

For the CL correlation we consider $k' \approx k$ and restrict ourselves to momenta $|k| \approx k_0$. In this case, the second line of Eq.~(\ref{eqn:ShortTime_G2_step2}) will not contribute (see Appendix \ref{app:ShortTime_g2} for details and later discussion of the BB correlation) and the CL correlation simplifies to: 
\begin{eqnarray}
 G^{(2)}_{\mathrm{CL}}(k,k',t) \simeq n(k,t)n(k',t) + |n(k,k',t)|^2 . \label{eqn:G2_CL_ShortTime}
\end{eqnarray}

Using Eqs.~(\ref{eqn:n_kkdash_ShortTime}) and (\ref{eqn:n_kk_ShortTime}) the normalized CL correlation is then given by:
\begin{equation}
 g^{(2)}_{\mathrm{CL}}(k,k',t) \simeq 1 + \left|\frac{\Gamma(7/2)J_{5/2}(\Delta{k}R_x)}{(\Delta{k}R_x/2)^{5/2}}\right|^2 , \label{eqn:g2_CL_ShortTime}
\end{equation}
where $\Delta k=k'-k$. Again, this result is unchanged to that of a true 1D condensate source in the undepleted pump approximation \cite{Ogren_Correlations_2009} and so phase-fluctuations play no role in the CL correlation as the phase operators cancel each other. Specifically, the peak (above unity) of the normalized correlation $h_{\mathrm{CL}}$ and the half-width at half-maximum (HWHM) $w_{\mathrm{CL}}$ of the CL correlation (above the background value of unity) are, respectively,
\begin{eqnarray}
 h_{\mathrm{CL}} & \equiv & g^{(2)}_{\mathrm{CL}}(k,k) - 1 = 1 , \\
 w_{\mathrm{CL}} & \simeq & \sqrt{\frac{7}{2}}\frac{1}{R_x} . \label{eqn:w_CL_ShortTime}
\end{eqnarray}
Both results are consistent with the observations of Ref.~\cite{KK_MultimodeCS_2012}. In particular, the $1/R_x$ scaling supports the observation that the width of the collinear correlation is related to the spatial density profile of the source cloud \cite{Ogren_Correlations_2009}, whilst the peak $h_{\mathrm{CL}}$ is driven by the Hanbury-Brown-Twiss effect \cite{Schellekens2005}, with $g^{(2)}_{\mathrm{CL}}(k,k,t) = 2$.

For the BB correlation, on the other hand, we have that $k' \approx -k$ and we again consider $|k| \approx k_0$. Under these conditions the correlation $n(k,k',t)$ is negligibly small and hence the BB correlation reduces to:
\begin{eqnarray}
& & G^{(2)}_{\mathrm{BB}}(k,k',t) \simeq n(k,t)n(k',t) \notag\\
& & + \left(\frac{g_{01}t}{2\pi\hbar}\right)^2  \iint dx dx' ~ e^{-i(k+k')(x-x')} \rho(x)\rho(x') e^{-2\langle (\delta\phi_{xx'})^2 \rangle} ,  \notag\\
\label{eqn:G2_BB_ShortTime_PhaseFluct}
\end{eqnarray}
for  $k^{\prime} \approx -k$
In contrast to the previous result for the CL correlation, the second line of Eq.~(\ref{eqn:G2_BB_ShortTime_PhaseFluct}) indicates that this correlation is sensitive to the phase-fluctuations of the source. 

An exact evaluation of the integral in Eq.~(\ref{eqn:G2_BB_ShortTime_PhaseFluct}) is not possible, however, by introducing the aforementioned approximation $\langle (\delta\phi_{xx'})^2 \rangle \simeq |x-x'|/l_T$ 
\cite{Petrov_1DqBEC_2000} and considering sufficiently high temperatures such that $l_T \ll R_x$ the integral may be evaluated (see Appendix \ref{app:ShortTime_g2} for details). We find that the normalized correlation is Lorentzian,
\begin{equation}
 g^{(2)}_{\mathrm{BB}}(k,k',t) \simeq 1 + \frac{h_{\mathrm{BB}}}{1 + \left( \Delta{k}/w_{\mathrm{BB}} \right)^2 } , \label{eqn:g2_BB_ShortTime}
\end{equation}
where $\Delta k=k'+k$, and where the peak height and HWHM of the correlation function are given, respectively, by
 \begin{eqnarray}
 h_{\mathrm{BB}} & = & \frac{4}{15\pi} \left[ \frac{\Gamma(7/2)\hbar}{g_{01}\rho(0)t} \right]^2 \frac{ l_T}{R_x} , \label{eqn:h_BB_ShortTime} \\
 w_{\mathrm{BB}} & = & \frac{2}{l_T} . \label{eqn:w_BB_ShortTime}
\end{eqnarray}

Equations~(\ref{eqn:g2_BB_ShortTime})-(\ref{eqn:w_BB_ShortTime}) are the key result of the short-time analytic model. 
Insight can be gained by contrasting the HWHM of the BB correlation against the momentum width of the source quasi-condensate $w_{k} \sim 1/l_T$. The result $w_{\mathrm{BB}} \sim w_{k}$ explicitly illustrates the expectation that the width of the BB correlation is proportional to the momentum width of the source, agreeing with previous discussions in Ref.~\cite{KK_MultimodeCS_2012} pertaining to the related process of collisions of elongated 3D quasi-condensates. Moreover, this result remains consistent with the result for a $T=0$ true condensate reported in Ref.~\cite{Ogren_Correlations_2009}, wherein $w_{\mathrm{BB}} \sim 1/R_x \sim w_k$. Lastly, we highlight that Eqs.~(\ref{eqn:g2_BB_ShortTime})-(\ref{eqn:w_BB_ShortTime}) predict that the net (integrated) correlation of the scattered pairs is preserved as the effective measure of the area under the correlation curve, $h_{\mathrm{BB}}w_{\mathrm{BB}}$, remains constant independent of temperature.

\section{Numerical simulation of twin beam production 
\label{sec:NumericalSims}}

As a benchmark of the analytic theory we compare the predictions to numerical simulations based on the stochastic positive-$P$ phase-space representation. In contrast to previous work \cite{Krachmalnicoff2010,KK_MultimodeCS_2012,RLS_HOM_2014,RLS_Bell_2015,Deuar_2011_Bog,Deuar_NumberSqueeze_2013}, 
our stochastic simulations do not depend on the implementation of a Bogoliubov linearization scheme for the phase-space variables. Moreover, unlike Sec.~\ref{sec:ShortTimeTheory} we do not invoke any undepleted pump approximation for the quasi-condensate source. Instead we model the dynamics of the complete system governed by Eqs.~(\ref{eqn:H0}) and (\ref{eqn:Hamiltonian}) using full positive-$P$ method as in Refs.~\cite{Deuar_2007,Perrin_AtomicFWM_2008,Midgley_2009,Savage_2006,Savage_2007}, where stochastic averages 
of the products of phase-space variables correspond, up to high-energy momentum cut-off errors in the modelling of the initial quasi-condensate [specifically, the phase-space representation of the phase operator $\hat{\phi}(x)$], to the exact quantum mechanical expectation values in the limit of an infinite number of trajectories. We direct the interested reader to Appendix~\ref{app:PosP_eqns} for the full details of the numerical model including the treatment of the phase operator.

In our simulation example, we model the twin beams produced from a quasi-condensate source as reported in the experiment of Ref.~\cite{Bucker_TwinBeams_2011}. Specifically, we model a phase fluctuating quasi-condensate of approximately 700 $^{87}$Rb atoms and ignore density fluctuations. We do not model the initial preparation of the quasi-condensate in the excited state, which is achieved by parametrically shaking the trap \cite{Bucker_OptimalControl_2013}, and instead assume that it is initially transferred to the $n_y = 1$ trap level with perfect fidelity. The confining trap is modelled as a harmonic potential with frequencies $(\omega_x,\omega_y,\omega_z)/2\pi = (16,1830,2580)$~Hz.

\subsection{Twin beams}
\begin{figure}
\centering
\includegraphics[width=8.0cm]{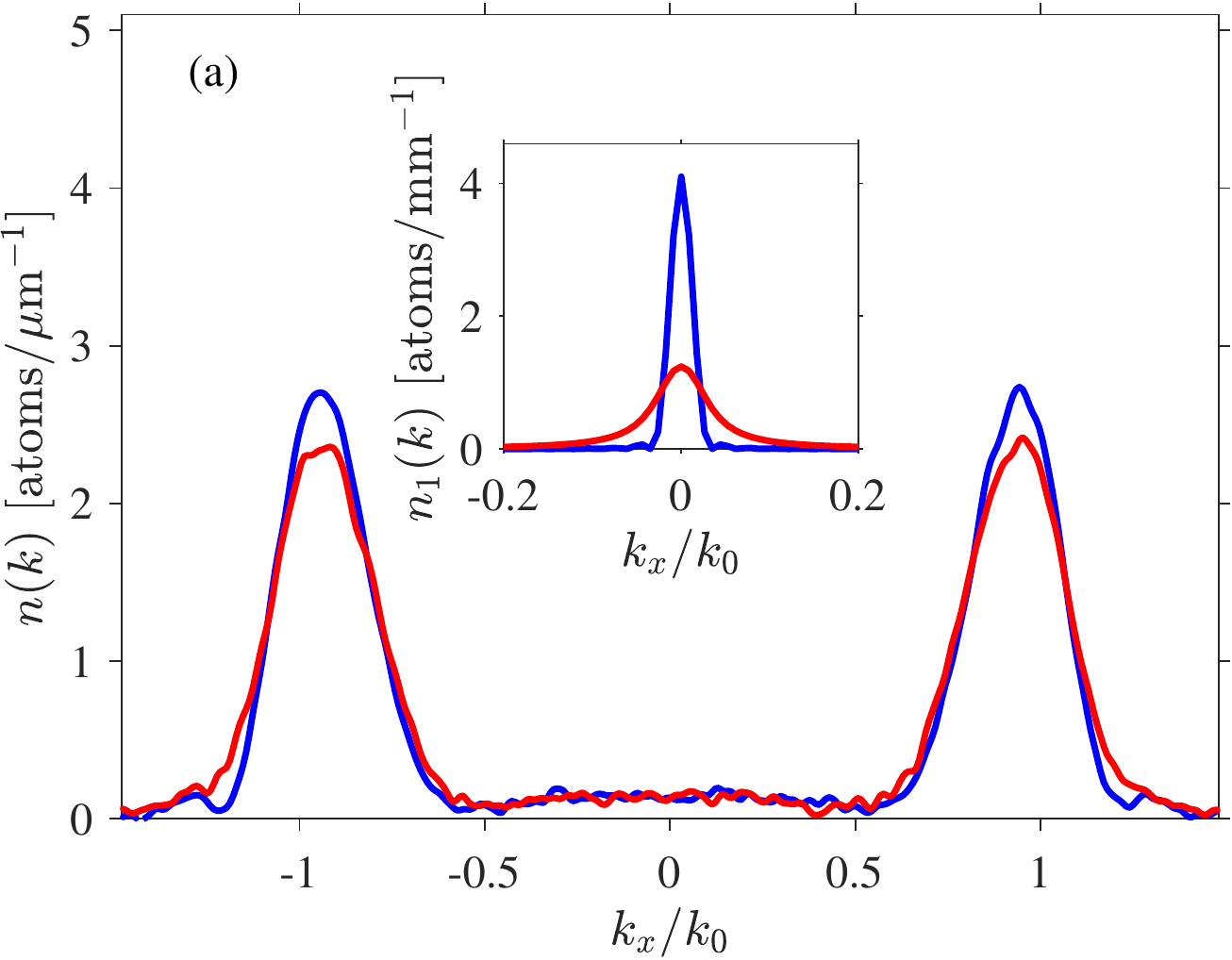}
\includegraphics[width=8.0cm]{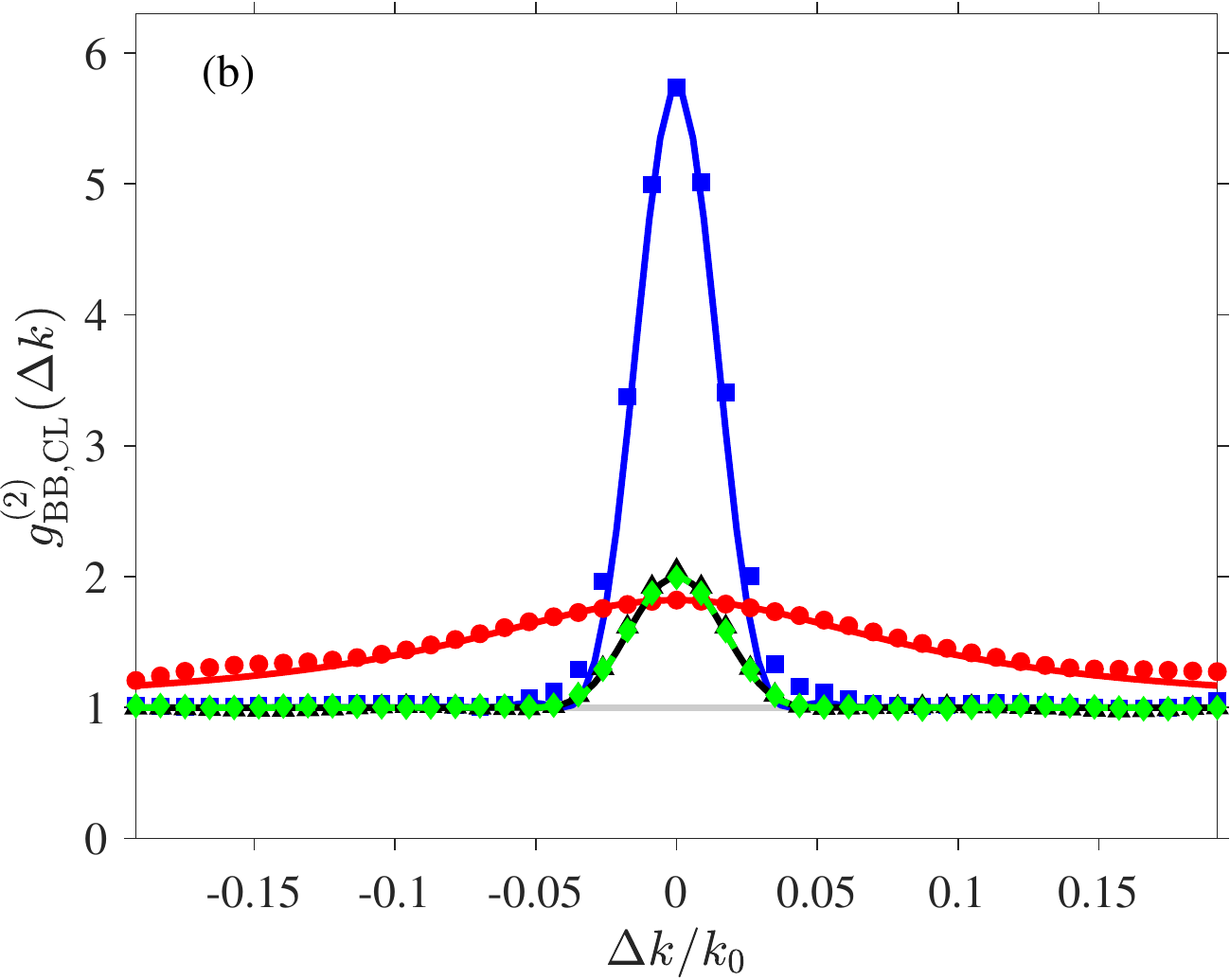}
\caption{(a) Momentum-space density profile $n(k)$ of the scattered atoms at $t=0.48$~ms for $T=0$~nK (blue line) and $T=40$~nK (red line). In the inset we plot the initial momentum-space 
density profiles of the source, $n_1(k)$, for $T=0$~nK (true BEC, blue line) and $T=40$~nK (quasi-condensate, red line). (b) Example of second-order correlation functions, $g^{(2)}_{\mathrm{BB},\mathrm{CL}}(\Delta k)$, at $t=0.35$~ms. We compare the BB correlation 
for a $T=0$~nK true BEC (blue squares) and $T=40$~nK quasi-condensate (red circles) to demonstrate the broadening and reduction of the correlation due to phase-fluctuations. The CL correlation is unchanged for $T=0$~nK (green diamonds) and 
$T=40$~nK (black triangles) thus not affected by the sources phase-fluctuations. The solid lines are fits based on Eqs.~(\ref{eqn:g2_BB_ShortTime}) and (\ref{eqn:g2_CL_ShortTime}), with widths given by the short-time analytic results (see text) and peak values re-scaled to numerical data for better comparison of the functional form. The straight grey line indicates the background (uncorrelated) level of $g^{(2)}(k,k') = 1$. }
\label{fig:Correlations_and_popn_dk} 
\end{figure}

In Fig.~\ref{fig:Correlations_and_popn_dk}~(a) we plot the results of the positive-$P$ calculation for: (\textit{i}) a pure ($T=0$~nK) condensate and (\textit{ii}) a $T=40$~nK quasi-condensate (the estimated temperature of the quasi-condensate of Ref.~\cite{Bucker_TwinBeams_2011}) 
at time $t=0.48$~ms~\footnote{The positive-$P$ calculations remain stable until $t\simeq0.65$~ms or $\simeq 2.5\%$ depletion.}.
(for comparison, $t_0\simeq0.5$~ms is time before which we expect the analytic model of the previous section to be valid). 
The twin beams are clearly identifiable and centered at $k \approx \pm 0.96k_0$, with the discrepancy from the exact value of $k=\pm k_0$ being due to the mean-field shift from the elastic scattering terms in Eq.~(\ref{eqn:Hamiltonian}) \cite{Wasak_BogRegions_2014}. 
The marginal broadening of the density profile of the scattered atoms in the quasi-condensate case is due to the increased momentum width of the source. In qualitative agreement with the invariance of Eq.~(\ref{eqn:n_kk_ShortTime}) of the short-time model with temperature, we observe that the total number of scattered atoms in the twin-beams is independent of temperature. 

\subsection{Atom-atom correlations}
For a detailed comparison to the qualitative predictions of the short-time analytic model we also extract the relevant second-order correlations from the positive-$P$ simulations. To reduce the sampling error from the stochastic simulations we integrate the correlation functions over a fixed region in momentum space and so define the averaged second-order correlation function \cite{Perrin_AtomicFWM_2008},
\begin{equation}
 \bar{g}^{(2)}_{\mathrm{BB}}(\Delta k) \equiv \frac{\int_{\Lambda} dk ~ G^{(2)}_{\mathrm{BB}}(k,-k+\Delta k) }{ \int_{\Lambda} dk ~ n(k) n(-k+ \Delta k)} \label{eqn:g2_ave_defn}
\end{equation}
and similarly for $\bar{g}^{(2)}_{\mathrm{CL}}(\Delta k)$. Here, the relevant region of integration in momentum space, $\Lambda$, is centered on one of the twin beams, e.g., at $k=k_0$, with the width taken to correspond to the characteristic extent $\sim 0.5k_0$ of the highly occupied region of the twin beam, such that we integrate over $k_0 - 0.5k_0 < k < k_0 + 0.5k_0$. The results are plotted in Fig.~\ref{fig:Correlations_and_popn_dk}~(b) for the same initial sources as (a) and $t=0.35$~ms. As predicted by the analytic theory, the strength and width of the CL correlation 
is unaffected by the phase-fluctuations and resultant broadening of the source momentum distribution. In contrast, we find the BB correlation is substantially suppressed and broadened in the case of a quasi-condensate, as expected from the analytic model. For comparison to the predictions of the short-time analytic model we plot Eqs.~(\ref{eqn:g2_CL_ShortTime}) and (\ref{eqn:g2_BB_ShortTime}) with widths as predicted by Eqs.~(\ref{eqn:w_CL_ShortTime}) and (\ref{eqn:w_BB_ShortTime}) respectively (see Appendix~\ref{app:AppT0_g2} for the corresponding $T=0$ expressions). However, as the $t=0.35$~ms is close to the cutoff time $t_0 \sim 0.5$~ms for the validity of the short-time analytic model we do not expect the correlation strengths to quantitatively match. Consequently, we artificially fix the peak height to match those of the positive-$P$ results and first focus on the correlation widths predicted by the two approaches. We find excellent agreement with the functional forms predicted by the analytic model, despite not satisfying the condition $t \ll t_0$ for these results, suggesting that the short-time predictions remain a useful qualitative guide with respect to the correlation widths, even beyond their explicit regime of validity.

\begin{figure}
\centering
\includegraphics[width=8.0cm]{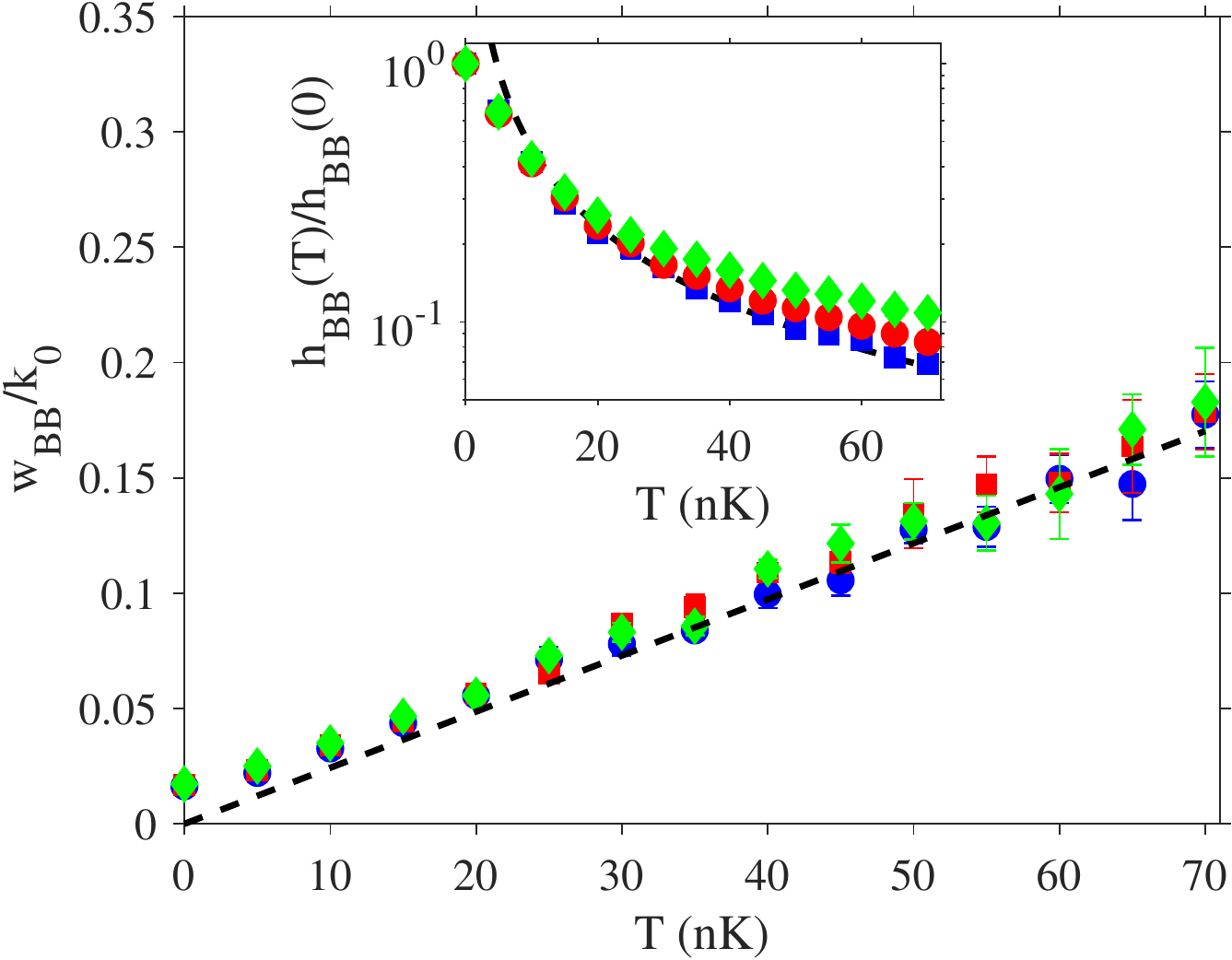}
\caption{Properties of BB correlation as a function of quasi-condensate initial temperature at $t=0.05$~ms (blue circles), $t=0.175$ms (red squares) and $t=0.3$~ms (green diamonds). We compare the correlation width $w_{\mathrm{BB}}$ (main plot) extracted from numerical simulations to the prediction of the short-time model [Eq.~(\ref{eqn:w_BB_ShortTime}), dashed black line]. Also plotted in the inset is the normalized peak correlation strength $h_{\mathrm{BB}}(T)/h_{\mathrm{BB}}(0)$ calculated from 
numerical simulations, compared to a $1/T$ fit (dashed black line) motivated by the analytic prediction of Eq.~(\ref{eqn:h_BB_ShortTime}). In both plots we find excellent (qualitative) agreement 
for $t=0.05$~ms and $T\gtrsim15$~nK. For longer times the peak correlation strength no longer follows the $1/T$ scaling, however, the correlation width remains relatively unchanged. 
}
\label{fig:Corr_T}
\end{figure}

To gain further insight, we compare the predicted scaling of correlation strengths and widths with temperature in Fig.~\ref{fig:Corr_T} for a range of time: $t=0.05$~ms, $t=0.175$~ms and $t=0.3$~ms. We find excellent quantitative agreement between the short-time analytic and positive-$P$ numerical results for the widths of the BB correlation for $T\gtrsim15$~nK and across the range of time samples. 
We observe clear linear scaling with $T$, agreeing with the short-time analytic prediction of Eq.~(\ref{eqn:w_BB_ShortTime}) [and given that $l_T = \hbar^2\rho_0/(m k_B T)$]. The minor disagreement for $T \lesssim 15$~nK is attributable to the poor fulfilment of the condition $l_T \gg R_x$ used in the derivation of Eq.~(\ref{eqn:g2_BB_ShortTime}) (see Appendix~\ref{app:ShortTime_g2}).
In the inset of Fig.~\ref{fig:Corr_T} we plot the peak BB correlation strength, 
normalized by the $T=0$ result. This normalization is used as the results of the positive-$P$ calculations do not quantitatively match the short-time model for explicit values of $h_{\mathrm{BB}}(T)$ -- due to the combination of 
the inhomogeneity of the twin-beams compared to the simple uniform scattering predicted by Eq.~(\ref{eqn:n_kk_ShortTime}) (which is only a reasonable characterisation for extremely short times $t \ll t_0$), and the 
use of averaged correlation functions [Eq.~(\ref{eqn:g2_ave_defn})]. For times up to $t\simeq 0.3$~ms, we find good agreement with the analytic prediction $h_{\mathrm{BB}}(T) \propto 1/T$ of Eq.~(\ref{eqn:h_BB_ShortTime}) for $T\gtrsim15$~nK.

\section{Applications to tests of Bell's inequality \label{sec:BellTest}}
A potential application for the twin-beams produced by collisional de-excitation is in tests of entanglement \cite{Bonneau_DoubleWell_2017} and fundamental tests of quantum mechanics, specifically a demonstration of a violation of a Bell inequality with massive particles using motional degrees of freedom \cite{RLS_Bell_2015,Dussarrat_Interferometer_2017}. Such a scheme has previously been studied in detail by the authors in Ref.~\cite{RLS_Bell_2015}, utilizing the related process of twin-atom production 
via collisions of pure phase-coherent Bose-Einstein condensates in combination with a Rarity-Tapster interferometric scheme. However, a crucial question is whether using a phase fluctuating source to produce the correlated atom-pairs will fundamentally affect any violation of the inequality. Specifically, a test of a Bell inequality generically requires measuring and characterising phase-sensitive correlations, which may be altered by phase-fluctuations in the 1D source. In this section we investigate this question in detail, using the extensive characterisation of the $G^{(2)}$ correlation functions in Sec.~\ref{sec:ShortTimeTheory} to understand how any possible violation scales with the phase coherence length and thus temperature of the quasi-condensate.


\begin{figure}
\centering
\includegraphics[width=8.8cm]{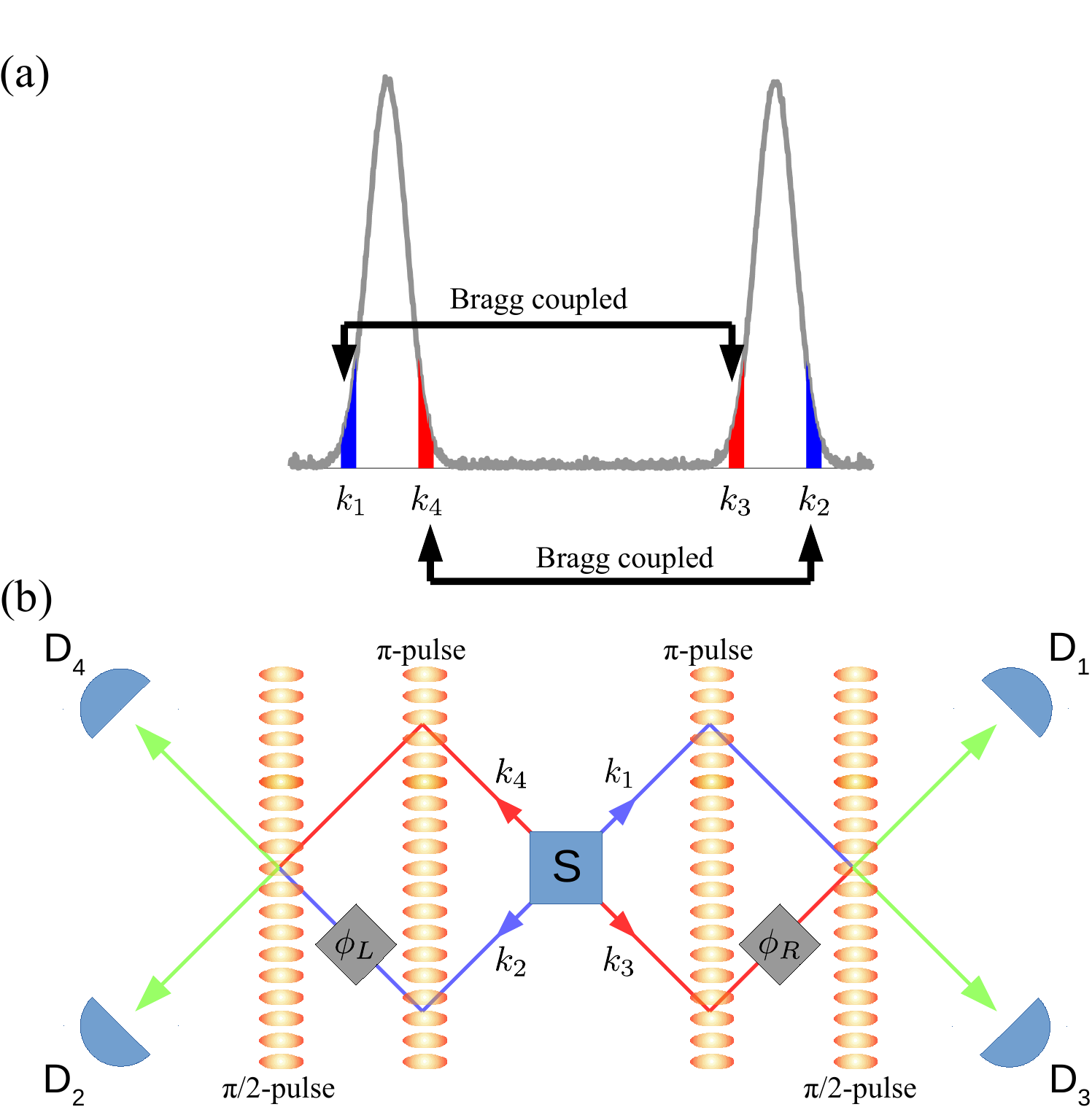}
\caption{(a) Illustration of twin beams with (example) targetted pairs of momenta $(k_1,k_2)$ (indicated in blue) and $(k_3,k_4)$ (indicated in red). Arrows indicate which momenta are coupled by the Bragg pulses. 
(b) Illustrative schematic of the Rarity-Tapster interferometer. The multimode source (S) correlated atom pairs in the twin-atom beams which travel through the left and right arms of the atomic interferometer. Application of a standing-wave 
light field realizes a $\pi$ Bragg pulse which couples momenta $k_1$ and $k_3$ ($k_2$ and $k_4$), before a phase-shift $\phi_R$ ($\phi_L$) is applied to the momentum component $k_3$ ($k_2$). After a period of free 
propagation (such that the atomic wavefpackets overlap in position space) a $\pi/2$ Bragg pulse is applied, coupling the same momenta. Atom-atom correlations are then measured between detectors $D_i$ ($i=1,2,3,4$), 
corresponding to coincidence counts $C_{ij}$ (see main text) from which the CHSH-Bell parameter is then constructed.
}
\label{fig:BellScheme} 
\end{figure}

While we point the interested reader to Ref.~\cite{RLS_Bell_2015} for a detailed description of the atomic Rarity-Tapster scheme, we include a simple illustration of the protocol in Fig.~\ref{fig:BellScheme}. 
The protocol requires choosing two pairs of momentum-correlated components $(k_1,k_2)$ and $(k_3,k_4)$ from the scattered twin-beams. These are chosen so that $k_2 = -k_1$ and $k_4 = -k_3$ and thus each pair will have a significant BB correlation due to the pair-wise scattering process. However, we choose the pairs such that they are independent and uncorrelated: This requires $|k_1 - k_4| \gg w_{\mathrm{CL}}$ and $|k_2 - k_3| \gg w_{\mathrm{CL}}$ to avoid any undesired CL correlation between the pairs $(k_1,k_4)$ and $(k_2,k_3)$. Moreover, we also require $|k_1 + k_3| \gg w_{\mathrm{BB}}$ and $|k_2 + k_4| \gg w_{\mathrm{BB}}$ to avoid any residual BB correlation between $(k_1,k_3)$ and $(k_2,k_4)$. 

The Rarity-Tapster interferometric scheme is realized by coupling the modes $k_1$ and $k_3$ ($k_2$ and $k_4$) via a Bragg $\pi$-pulse (the atomic equivalent of an optical mirror) to reverse their motion in position space and bring the atomic wavepackets together such that they spatially overlap. During the ensuing free-propagation of the wavepackets (and before they recombine spatially), 
we imprint variable phase-shifts of $\phi_L$ and $\phi_R$ in the `lower-arm' of the interferometer, \textit{i.e.}, on modes $k_2$ and $k_3$. Upon recombination of spatially overlapping wavepackets, we mix the modes $k_1$ and $k_3$ ($k_2$ and $k_4$) with a Bragg $\pi/2$-pulse (atomic equivalent of a $50$-$50$ optical beam-splitter). Finally, we measure atom-atom cross-correlations between the modes at the time designated $t_f$.
 
A Bell inequality is constructed from this interferometric protocol by measurement of  a set of atom-atom correlations $C_{ij} \equiv G^{(2)}(k_i,k_j,t_f)$ at time $t_f$ (the output of the interferometer) between detectors $D_i$ and $D_j$ (see Fig.~\ref{fig:BellScheme}) for specific choices of the phase settings $(\phi_L,\phi_R)$. From these, one constructs the (phase-sensitive) correlation coefficient \cite{CHSH_Bell_1969,Aspect_Bell_1982}
\begin{equation}
 E(\phi_L,\phi_R) \equiv \left.\frac{C_{14} + C_{23} - C_{12} - C_{34}}{C_{14} + C_{23} + C_{12} + C_{34}}\right\vert_{\phi_L,\phi_R} . \label{eqn:E_phi}
\end{equation}
The CHSH-Bell parameter $S$ is then defined as \cite{CHSH_Bell_1969}:
\begin{equation}
 S = \vert E(\phi_L,\phi_R) + E(\phi_L,\phi^{\prime}_R) + E(\phi^{\prime}_L,\phi_R) + E(\phi^{\prime}_L,\phi^{\prime}_R) \vert .
\end{equation}
The CHSH-Bell inequality then dictates that any local hidden-variable theory of quantum mechanics must satisfy $S\leq2$ \cite{CHSH_Bell_1969}. However, certain states in quantum mechanics, such as the idealized twin-atom state \cite{RLS_Bell_2015}, are known to be strongly correlated such that $S > 2$, and thus they are said to violate the CHSH form of a Bell inequality. The twin-atom state, in particular, maximally violates the CHSH-Bell inequality by saturating the quantum bound $S = 2\sqrt{2}$ \cite{Tsirelson1980}.

The dependence of the atom-atom correlation functions $C_{ij}$ and thus the correlation coefficient $E(\phi_L,\phi_R)$ on the phase-settings $\phi_L, \phi_R$ is crucial to the CHSH-Bell inequality. A priori, one might be concerned as to whether the phase-fluctuations of the source destroy any possibility of a violation, primarily because the sensitivity to the phase-settings $\phi_{L,R}$ might be destroyed if the scattered pairs ($k_1, k_2$) and ($k_3, k_4$) do not have a well-defined phase relationship. To elucidate this statement and gain some simple intuition into the role of the phase-fluctuations of the quasi-condensate we make a momentary diversion in order to develop a toy-model that clarifies the question at hand.

\subsection{Toy model}
Our toy model begins by considering a simplified description of the collisional de-excitation process, ignoring spatial structure and elastic collisions, and is described by the few-mode Hamiltonian:
\begin{equation}
 \hat{H} = \hbar g \left( \hat{b}^{\dagger}_A\hat{b}^{\dagger}_A\hat{a}_1\hat{a}_2 + \hat{b}^{\dagger}_B\hat{b}^{\dagger}_B\hat{a}_3\hat{a}_4 + h.c. \right) .
\end{equation}
This Hamiltonian can be considered as a simplification of the four-wave mixing Hamiltonian $\hat{H}_{\mathrm{FWM}}$ [Eq.~(\ref{eqn:H_4WM})] in which we further assume the de-excitation process is restricted to scatter particles into only two pairs of momentum modes which we label $(1,2)$ and $(3,4)$ (rather than many momentum modes within the twin beams). Moreover, by assuming each pair of modes is populated from an independent source mode [$\hat{b}_A$ and $\hat{b}_B$], we are effectively considering the case where particles scattered from the quasi-condensate are created in distinct spatial regions~\footnote{This is a reasonable assumption if we consider only very weak scattering such that approximately only one or zero pairs are scattered, and the quasi-condensate is sufficiently hot so that $l_T$ is much smaller than the spatial extent of the quasi-condensate.}. To simplify the following analysis we  then invoke an undepleted pump approximation and replace the source mode bosonic operators with $c$-numbers, $\hat{b}_B \to \beta$ and $\hat{b}_A \to \beta e^{i\varphi}$. The relative phase difference $\varphi$ encapsulates the fact that differing spatial regions in the quasi-condensate will not be phase-coherent (in contrast to a BEC which possesses true long-range order). 

Assuming an initial vacuum condition for the modes $(1,2)$ and $(3,4)$, then in the limit of very weak scattering $g\vert\beta\vert^2t \ll 1$ the output state of this model in the Schr\"{o}dinger picture can be approximated in the Fock basis as \cite{RLS_HOM_2014,Braunstein_RevModPhys_2005}:
\begin{multline}
 \vert \psi(t) \rangle \approx \vert 0_1, 0_2, 0_3, 0_4 \rangle \\
 - ig\beta^2t\left( e^{2i\varphi}\vert 1_1, 1_2, 0_3, 0_4 \rangle + \vert 0_1, 0_2, 1_3, 1_4 \rangle \right) . \label{eqn:SimpleState}
\end{multline}
The average mode occupation here is $n \equiv \langle \hat{a}^{\dagger}_j\hat{a}_j \rangle \approx g^2\vert \beta \vert^2 t^2$ for $j=1,2,3,4$. 

Treating the Bragg $\pi$ and $\pi/2$-pulses in the Rarity-Tapster interferometric scheme as a sequence of linear transformations  \cite{RLS_Bell_2015} we can evaluate the atom-atom correlations $C_{ij}$ with respect to the state Eq.~(\ref{eqn:SimpleState}). As an example let us consider $C_{12}$ which can be expressed as:
\begin{multline}
 C_{12} = \frac{1}{4} \left[ \langle \hat{a}^{\dagger}_1 \hat{a}^{\dagger}_2 \hat{a}_2 \hat{a}_1 \rangle + \langle \hat{a}^{\dagger}_3 \hat{a}^{\dagger}_4 \hat{a}_4 \hat{a}_3 \rangle  \right. \\
 + \langle \hat{a}^{\dagger}_1\hat{a}^{\dagger}_4\hat{a}_4\hat{a}_1\rangle + \langle \hat{a}^{\dagger}_2\hat{a}^{\dagger}_3\hat{a}_3\hat{a}_2\rangle \\
 + \langle \hat{a}^{\dagger}_4\hat{a}^{\dagger}_3 \hat{a}_2 \hat{a}_1 \rangle e^{i(\phi_L-\phi_R)} 
 + \left. \langle \hat{a}^{\dagger}_1\hat{a}^{\dagger}_2 \hat{a}_3 \hat{a}_4 \rangle e^{-i(\phi_L-\phi_R)} \right] . \label{eqn:BellPhaseCorr}
\end{multline}
Of note is that the phase-dependence of the atom-atom correlations, which are key to the violation of the CHSH-Bell inequality, stem from the last two lines which involve interference of the two scattered pairs. 
Evaluation of each contributing term with respect to the state Eq.~(\ref{eqn:SimpleState}) leads to
\begin{eqnarray}
 \langle \hat{a}^{\dagger}_1 \hat{a}^{\dagger}_2 \hat{a}_2 \hat{a}_1 \rangle & = & \langle \hat{a}^{\dagger}_3 \hat{a}^{\dagger}_4 \hat{a}_4 \hat{a}_3 \rangle = g^2\beta^4t^2 , \\
 \langle \hat{a}^{\dagger}_1 \hat{a}^{\dagger}_2 \hat{a}_2 \hat{a}_1 \rangle & = & \langle \hat{a}^{\dagger}_3 \hat{a}^{\dagger}_4 \hat{a}_4 \hat{a}_3 \rangle = 0 , \\
 \langle \hat{a}^{\dagger}_4 \hat{a}^{\dagger}_3 \hat{a}_2 \hat{a}_1 \rangle & = &  g^2\beta^4t^2 e^{2i\varphi} , \\
 \langle \hat{a}^{\dagger}_1 \hat{a}^{\dagger}_2 \hat{a}_3 \hat{a}_4 \rangle & = & g^2\beta^4t^2 e^{-2i\varphi} ,
\end{eqnarray}
which finally yields:
\begin{equation}
 E(\phi_L, \phi_R) \approx \frac{1}{2}\mathrm{cos}(\phi_L - \phi_R + 2\varphi) .
\end{equation}
Here, we find the phase-difference of the source modes feeds directly into the phase-dependence of the correlation coefficient in the last two terms in Eq.~(\ref{eqn:BellPhaseCorr}). For a fixed phase-difference $\varphi$ the oscillating fringe of the correlation coefficient is simply shifted. In particular, the CHSH-Bell inequality may still be nearly maximally violated by appropriately shifting the optimal phase-settings, which for $\varphi=0$ are given by $(\phi_L,\phi_R,\phi^{\prime}_L,\phi^{\prime}_R) = (0, \pi/4, \pi/2, 3\pi/4)$ and yield $S = 2\sqrt{2}$. 

On the other hand, treating the phase-difference as stemming from the thermal phase-fluctuations of a quasi-condensate leads to $\varphi$ being best described as a Gaussian random variable from shot-to-shot. Evaluation of the correlation functions $C_{ij}$ then leads to an exponential suppression of the correlation coefficient,
\begin{equation}
 \langle E(\phi_L, \phi_R) \rangle_{\mathrm{fluct}} \approx \frac{e^{-2(\Delta\varphi)^2}}{2} \mathrm{cos}(\phi_L - \phi_R)
 \end{equation}
where $\langle ... \rangle_{\mathrm{fluct}}$ indicates stochastic averaging over the phase-fluctuations for which $\varphi$ is a real Gaussian random variable with zero mean and variance $(\Delta\varphi)^2$. Here, the decay of the correlation coefficient due to the phase-fluctuations implies that the CHSH-Bell parameter is also unavoidably suppressed as $S_{\mathrm{fluct}} = 2\sqrt{2}e^{-2(\Delta\varphi)^2}$, implying that for a violation to be preserved, $S_{\mathrm{fluct}} > 2$, the phase-fluctuations must be limited to $(\Delta\varphi)^2 < (1/2)\mathrm{log}(\sqrt{2})$.

This simplified toy model then captures the essential effect of phase-fluctuations in an intuitive way: The phase-sensitive correlations, which are required to demonstrate a violation of the CHSH-Bell inequality, are washed out when the scattered pairs do not possess a well defined phase relationship, i.e. when they are produced by a source which lacks phase-coherence. We can then expect that if the pairs are produced by a quasi-condensate with large phase-fluctuations the possible violation will be markedly reduced if not completely destroyed. Building on this qualitative result, in the next section we present a more detailed semi-analytic calculation based on the short-time analytic formalism of Sec.~\ref{sec:ShortTimeTheory} in combination with positive-$P$ simulations.

\subsection{Numerical model}
As a better approximation to experimentally realistic systems, we first generalize our analysis by introducing integrated correlation functions (to account for finite experimental detector resolution) $\mathcal{C}_{ij} = \langle:\hat{N}_i\hat{N}_j:\rangle$ for $i,j=1,2,3,4$, where $\hat{N}_i \equiv \int_{\Lambda_i} \hat{n}(k) dk$ is the integrated momentum-space atomic density in the region $\Lambda_i$ and the double columns stand for normal ordering of the respective creation and annihilation operators, $\hat{a}^{\dagger}(k)$ and $\hat{a}(k)$. Substitution of $\mathcal{C}_{ij}$, which can be parametrized in terms of the amplitudes and widths of the BB and CL correlations (see Appendix~\ref{app:BellDeriv} for details), into the correlation coefficient $\mathscr{E}$, as in Eq.~(\ref{eqn:E_phi}), allows us to then define a generalized multimode CHSH-Bell parameter $\mathcal{S}$ \cite{RLS_Bell_2015}:
\begin{equation}
 \mathcal{S} = 2\sqrt{2} \frac{h_{\mathrm{BB}} \mathcal{B}_{\mathrm{BB}}}{\frac{\mathcal{K}^2}{2} + h_{\mathrm{BB}} \mathcal{B}_{\mathrm{BB}}} , \label{eqn:IntegratedBell}
\end{equation}
where
\begin{multline}
 \mathcal{B}_{\mathrm{BB}} =  2 \mathcal{K} w_{\mathrm{BB}}\, \mathrm{atan} \left(\frac{\mathcal{K}}{4w_{\mathrm{BB}}}\right) \\ - 4w_{\mathrm{BB}}^2 \mathrm{log}\left[ 1 + \left( \frac{\mathcal{K}}{4w_{\mathrm{BB}}} \right)^2 \right], \label{eqn:B_BB}
\end{multline}
and $\mathcal{K}$ is the width of each integration region $\Lambda_i$ in momentum space: $k_i - \mathcal{K}/2 \leq k \leq k_i + \mathcal{K}/2$. We note that the form of Eq.~(\ref{eqn:IntegratedBell}) is subtilely different to that of Ref.~\cite{RLS_Bell_2015} due to the assumption of a Lorentzian rather than Gaussian form for the BB correlations. Violation of a Bell inequality again requires $\mathcal{S} > 2$, whilst the quantum bound remains $\mathcal{S} \leq 2\sqrt{2}$.

\begin{figure}
\centering
\includegraphics[width=8.5cm]{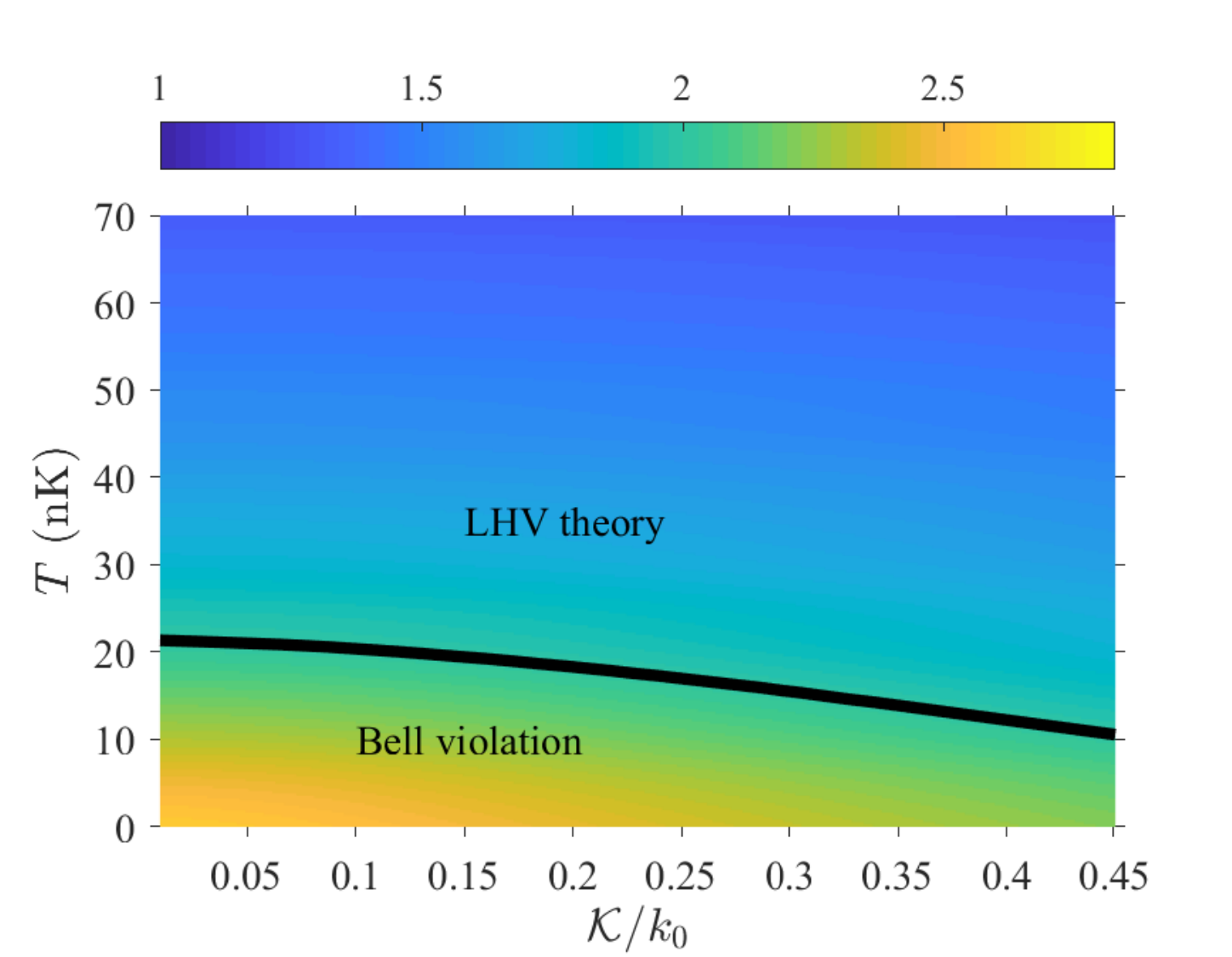}\\
\includegraphics[width=8.5cm]{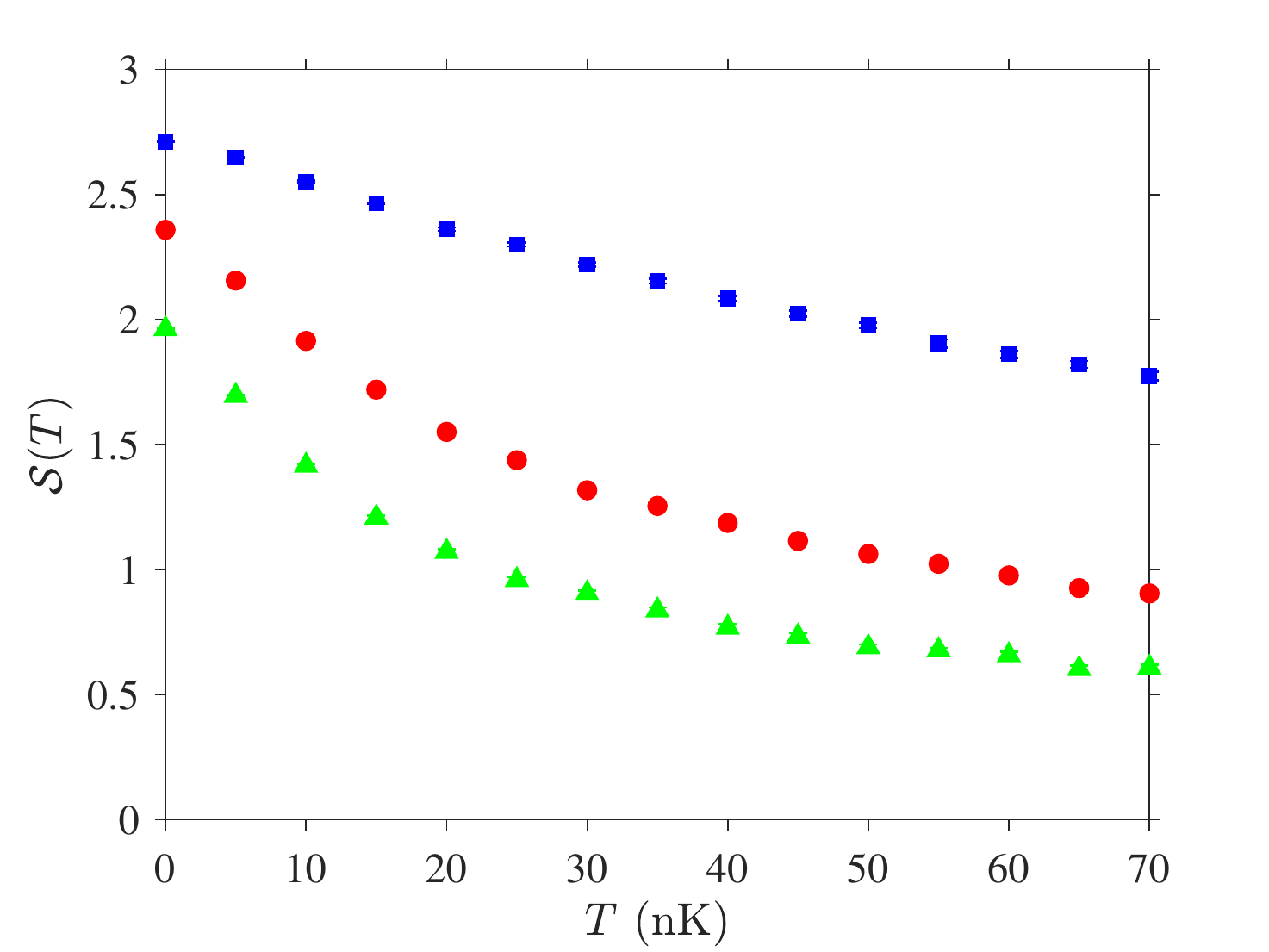}
\caption{(a) CHSH-Bell $\mathcal{S}$ parameter as a function of temperature $T$ and width $\mathcal{K}$ of integration region at fixed de-excitation duration $t=0.15ms$. Results are for Eq.~(\ref{eqn:IntegratedBell}) evaluated with 
$h_{\mathrm{BB}}$ and $w_{\mathrm{BB}}$ extracted from positive-$P$ simulations. Black line denotes crossover from a violation of the CHSH-Bell inequality ($\mathcal{S} > 2$) to results consistent with a local hidden variable (LHV) theory ($\mathcal{S} \leq 2$). (b) CHSH-Bell 
$\mathcal{S}$ parameter as a function of temperature with fixed integration width $\mathcal{K} \equiv \mathcal{K}(T) = 2w_{\mathrm{BB}}$. Similar to (a), results are for Eq.~(\ref{eqn:IntegratedBell}) evaluated with $h_{\mathrm{BB}}$ and $w_{\mathrm{BB}}$ extracted from 
positive-$P$ simulations for $t=0.1$~ms (blue squares), $t=0.23$~ms (red circles) and $t=0.35$~ms (green triangles). 
}
\label{fig:CHSH_T} 
\end{figure}

We investigate the behaviour of Eq.~(\ref{eqn:IntegratedBell}) with temperature $T$ (and thus implicitly the thermal phase coherence length) by substituting values of $h_{\mathrm{BB}}$ and $w_{\mathrm{BB}}$ extracted from positive-$P$ calculations. Parameters are chosen the same as those listed in Sec.~\ref{sec:NumericalSims}. In Fig.~\ref{fig:CHSH_T}~(a) we plot our results as a function of temperature $T$ and integration width $l$ at a fixed duration $t=0.23$~ms.

The rapid decay of $\mathcal{S}$ with increasing $T$ can be attributed to the suppression of the phase-sensitive correlations in Eq.~(\ref{eqn:BellPhaseCorr}) due to the increased phase-fluctuations of the quasi-condensate source with temperature, parametrized here by the dependence on $h_{\mathrm{BB}}\propto 1/T$ in the top line of Eq.~(\ref{eqn:IntegratedBell}). Moreover, $\mathcal{S}$ is found to decay as the integration region is increased. Whilst this has been discussed previously in Ref.~\cite{RLS_Bell_2015} in the context of a phase-coherent BEC source, we highlight that it contrasts to prior tests of non-classical phase-insensitive correlations, such as the Cauchy-Schwarz inequality \cite{KK_MultimodeCS_2012}, with a quasi-condensate source. The Cauchy-Schwarz inequality violation is highly dependent on the strength of the BB correlation, and it was found that the suppression of the correlation amplitude $h_{\mathrm{BB}}\sim1/T$ was offset by taking into account the broadening of the correlation function $w_{\mathrm{BB}} \propto T$ as the `net BB correlation' $h_{\mathrm{BB}}w_{\mathrm{BB}} = \mathrm{const.}$ was insensitive to $T$. However, in the context of the Bell inequality increasing the size $\mathcal{K}$ of the integration region is dominated by the increased number of uncorrelated (in the sense of phase-sensitive correlations) atoms in the detection region and thus does not increase $\mathcal{S}$.

The issue of introducing additional phase-uncorrelated particles into the integration region is illustrated in a complementary manner in Fig.~\ref{fig:CHSH_T}~(b), wherein we fix the size of the integration region to match the temperature dependent correlation width, $\mathcal{K} \equiv \mathcal{K}(T) = 2w_{\mathrm{BB}}$ such that the `net BB correlation' $\propto h_{\mathrm{BB}}w_{\mathrm{BB}}$ is preserved. Here, although $\mathcal{K}(T)$ now increases with temperature to compensate for the drop in peak correlation strength $h_{\mathrm{BB}}$ such that 
$h_{\mathrm{BB}}\mathcal{B}_{\mathrm{BB}} \propto T$, the number of (phase) uncorrelated particles in the integration region now increases, which is captured via the term $\mathcal{K}^2/2 \propto T^2$ in the denominator of Eq.~(\ref{eqn:IntegratedBell}). 

Overall then, a demonstration of a CHSH-Bell inequality violation requires the preservation of strong phase-sensitive correlations. In the context of the pair production process this can be cast as requiring a high peak BB correlation strength achieved via either: (i) ensuring the quasi-condensate source posseses sufficient phase-coherence (by, e.g., cooling to low enough $T$), or (ii) by reducing the duration of the pair-production process \cite{RLS_Bell_2015}.

\section{Conclusion \label{sec:Conclusion}}
In summary, we have shown that the key effect of the phase-fluctuations of a 1D quasi-condensate source is to lead to broadening of the back-to-back pair correlation of the twin-atom state. This is most clearly identified 
from a short-time analytic theory, which demonstrates that the broadening of the momentum distribution of the finite-temperature quasi-condensate leads to a broadening of the BB correlation width between scattered atoms 
and an associated suppression of the peak BB correlation strength. Specifically, the width $w_{\mathrm{BB}}$ is found to scale linearly with temperature $T$. We have validated these results via comparison to numeric calculations using the positive-$P$ stochastic phase-space method for short-times, and we 
also demonstrated that the qualitative predictions of the analytic model remain valid for longer times beyond the direct regime of validity of the model. 
Whereas the details of our model are focussed on the collisional de-excitation of atom pairs from a 1D quasi-condensate, we expect the formalism to be broadly applicable to 1D quasi-condensate systems and may be easily adapted to accomodate quasi-condensation in 3D Bose gases \cite{KK_MultimodeCS_2012,Bonneau_CorrelatedBeams_2013}. Moreover, our treatment of the phase-fluctuations in the short-time analytic model demonstrates that the undepleted pump approximation can be adapted to systems which lack coherence, and only requires us to assume absence of density fluctuations. 

We have directly applied the insights gained from the analytic model to demonstrate the impact of using a phase-fluctuating 1D source for a proposed violation of a motional-state Bell inequality. Unlike other, previously demonstrated, 
measures of non-classicality such as number squeezing \cite{KK_SubPoiss_2010} and the Cauchy-Schwarz inequality \cite{KK_MultimodeCS_2012}, we find that the phase-fluctuations of the source have important implications that can degrade and eventually 
destroy any violation of a CHSH-Bell inequality. These insights will have a direct impact on future tests of Bell inequalities for motional degrees of freedom of massive particles based on the utilization of twin-atoms from similar pair-production processes.

\begin{acknowledgments}
We acknowledge fruitful discussions with J\"{o}rg Schmiedmayer, Marie Bonneau, and thank Ana Maria Rey and John Bohn for feedback on the manuscript. This work is supported by the Australian Research Council Discovery Project grant DP170101423 (K.V.K.)  and  JILA-PFC NSF Grant No. PHY-1734006 (R.J.L.-S.).
\end{acknowledgments}

\appendix

\section{Evaluation of BB correlation in short-time approximation \label{app:ShortTime_g2}}
In this appendix we outline the solution of the BB correlation in the short-time analytic theory. We begin from Eq.~(\ref{eqn:G2_BB_ShortTime_PhaseFluct}) in the main text. 
Adopting the approximation $\langle (\delta\phi_{xx'})^2 \rangle \simeq -|x-x'|/l_T$, we focus on the unsolved integral which, ignoring prefactors, is:
\begin{equation}
 I \equiv  \int^{R_x}_{-R_x} dx \int^{R_x}_{-R_x} dx' ~ e^{-i(k+k')(x-x')} \rho(x)\rho(x') e^{-\frac{2|x-x'|}{l_T}} . 
\end{equation}

Transforming to sum and difference co-ordinates, $u=(x+y)/(\sqrt{2}R_x)$ and $v=(x-y)/(\sqrt{2}R_x)$, the integral may be rewritten as
\begin{eqnarray}
 I & = & R^2_x \int_{0}^{\sqrt{2}} dv~ \Bigg\{  e^{-i\sqrt{2}R_x(k+k')v}  \notag \\
 & & \times \Big[ e^{-\frac{2\sqrt{2}R_x}{l_T} v} \int_{v-\sqrt{2}}^{-v + \sqrt{2}} du~ f(u,v) \notag \\
 & & + e^{\frac{2\sqrt{2}R_x}{l_T} v} \int_{-v-\sqrt{2}}^{v + \sqrt{2}} du~ f(u,v) \Big] \Bigg\} ,
\end{eqnarray}
where 
\begin{equation}
 f(u,v) = \rho_0^2 \left( 1 - \frac{(u-v)^2}{2} \right)\left( 1 - \frac{(u+v)^2}{2} \right) .
\end{equation}
is the transformed product of the Thomas-Fermi density profiles. 

The integral over $u$ can be evaluated to give
\begin{eqnarray}
 I & = &  \frac{8R^2_x\rho_0^2}{15} \int_{0}^{\sqrt{2}} dv~ \Big\{  e^{-\frac{2\sqrt{2}R_x}{l_T} v} \mathrm{cos}\left[\Delta{k}R_x\sqrt{2}v\right] \notag \\
 & & \times  \left( -v^5 + 10v^3 - 10\sqrt{2}v^2 + 4\sqrt{2} \right) \Big\} ,
\end{eqnarray}
where we have introduced $\Delta{k} \equiv k + k'$ for $k'\approx -k$. For sufficiently high temperature, such that the spatial phase coherence length is much smaller than 
the condensate size $l_T/R_x \ll 1$, the exponential decay of the first term in the integral means we may extend the upper limit of integration to infinity. The integral 
is subsequently evaluated to give
\begin{equation}
 I = \frac{2R^2_x\rho^2_0}{15}\left\{ \frac{l_T}{R_x}\frac{1}{1 + \left(\frac{\Delta{k}l_T}{2} \right)^2 } + \mathcal{O}\left( \left[\frac{l_T}{R_x} \right]^3 \right) \right\} ,
\end{equation}
where $l_T/R_x$ is a small parameter. 

Substitution of this result in Eq.~(\ref{eqn:G2_BB_ShortTime_PhaseFluct}) and subsequent normalization by the atomic density [Eq.~(\ref{eqn:n_kk_ShortTime})] then gives the BB correlation 
as per Eq.~(\ref{eqn:g2_BB_ShortTime}), which is valid up to small corrections of $\mathcal{O}([l_T/R_x]^3)$.

\section{BB correlation function for $T=0$ \label{app:AppT0_g2}}
For clarity we also present the short-time prediction for the BB correlation where the source is a true 1D condensate (\textit{i.e.}, $T=0$ and the above derivation is invalid). This was previously addressed in Ref.~\cite{Ogren_Correlations_2009} for a 3D system, however, 
it is trivial to generalize the results to 1D. The BB correlation in this case is
\begin{equation}
 g^{(2)}_{\mathrm{BB}}(k,k',t) \simeq 1 + h^{\mathrm{BEC}}_{\mathrm{BB}}\left| \frac{\Gamma(5/2)J_{3/2}(\Delta{k}R_x)}{ \left(\frac{\Delta{k}R_x}{2} \right)^{3/2} } \right|^2 , \label{eqn:g2_BB_ShortTime_T0}
\end{equation}
where,
\begin{eqnarray}
 h^{\mathrm{BEC}}_{\mathrm{BB}} = \left(\frac{5\hbar}{2\Gamma(5/2)g_{01}\rho_0t}\right)^2 , \label{eqn:h_BB_ShortTime_T0}
\end{eqnarray}
and $\Delta{k} = k+k'$ as previous. The HWHM is found to be
\begin{eqnarray}
 w^{\mathrm{BEC}}_{\mathrm{BB}} \simeq \sqrt{\frac{5}{2}}\frac{1}{R_x} . \label{eqn:w_BB_ShortTime_T0}
\end{eqnarray}

\section{Integrated CHSH-Bell inequality \label{app:BellDeriv}}
As outlined in Sec. \ref{sec:BellTest}, a semi-analytic model of the CHSH-Bell quantity similar in form to Eq.~(\ref{eqn:IntegratedBell}) has previously been derived in Ref.~\cite{RLS_Bell_2015}. However, beyond the trivial reduction to 1D, there exist key differences for the case of a quasi-condensate source, particularly that Wick's theorem may not be applied to the expectation values of creation/annihilation operators, which limit its applicability to our scheme. In this appendix, we follow a similar procedure to Ref.~\cite{RLS_Bell_2015} and use physical insight gained via the short-time analytic model of Sec. \ref{sec:ShortTimeTheory} to derive Eq.~(\ref{eqn:IntegratedBell}) for a quasi-condensate source. 

To derive an expression for $\mathcal{S}$ we first consider the integrated pair-correlation functions after the application of the $\pi/2$ Bragg pulse (see Ref.~\cite{RLS_Bell_2015} for 
full details):
\begin{eqnarray}
 \mathcal{C}_{ij}(\phi_L,\phi_R) = \int^{k_i+\mathcal{K}/2}_{k_i-\mathcal{K}/2} dk \int^{k_j+\mathcal{K}/2}_{k_j-\mathcal{K}/2} dk' G^{(2)}(k,k',t_f) . \label{eqn:C_ij}
\end{eqnarray}
where $t_f$ is taken to be after the final $\pi/2$ Bragg pulse. For simplicity, we specialize to the case of $\mathcal{C}_{12}(\phi_L,\phi_R)$ in the following, with the generalization to the remaining correlations trivially accomplished.
Equation.~(\ref{eqn:C_ij}) can be written entirely in terms of correlations after the initial de-excitation (specified by $t=t_c$) and, treating the Bragg pulses as ideal mirrors and beam-splitters \cite{RLS_Bell_2015}, 
the integrand is given by:
\begin{widetext}
\begin{eqnarray}
 G^{(2)}(k,k',t_f) & = & \frac{1}{4} \Big[ G^{(2)}(k,k',t_c) + G^{(2)}(k+2k_L,k'-2k_L,t_c) + n(k,t_c)n(k'-2k_L,t_c) + n(k',t_c)n(k+2k_L,t_c) \notag \\
 & &  + \langle \hat{a}^{\dagger}(k+2k_L,t_c)\hat{a}^{\dagger}(k'-2k_L,t_c)\hat{a}(k,t_c)\hat{a}(k',t_c) \rangle e^{i(\phi_L - \phi_R)} \notag \\
 & & + \langle \hat{a}^{\dagger}(k,t_c)\hat{a}^{\dagger}(k',t_c)\hat{a}(k+2k_L,t_c)\hat{a}(k'-2k_L,t_c) \rangle e^{-i(\phi_L - \phi_R)} \Big] \label{eqn:G2_kkdash_t4}
\end{eqnarray}
\end{widetext}
where $2k_L = k_3 - k_1$ is the Bragg vector characterising the momentum transfer of the Bragg pulse. The first line can be readily simplified by making the replacement,
\begin{equation}
 G^{(2)}(k,k',t_c) \equiv \bar{n}^2 + \bar{n}^2\frac{h_{\mathrm{BB}}}{1 + \left(\frac{\Delta{k}}{w_{\mathrm{BB}}}\right)^2} , \label{eqn:G2_BellReplace}
\end{equation}
where $\Delta{k} \equiv k + k'$ as previous and we have assumed that the population in the relevant integration regions is approximately uniform with $n(k)\approx\bar{n}$.

For the case of a true condensate source, as in Ref.~\cite{RLS_Bell_2015}, the expectation values of the second and third lines could be simplified by a factorization 
using Wick's theorem. For a quasi-condensate, however, we use the short-time analytic model to show that
\begin{multline}
 \langle \hat{a}^{\dagger}(k+2k_L,t_c)\hat{a}^{\dagger}(k'-2k_L,t_c)\hat{a}(k,t_c)\hat{a}(k',t_c) \rangle \\
 \equiv \bar{n}^2\frac{h_{\mathrm{BB}}}{1 + \left(\frac{\Delta{k}}{w_{\mathrm{BB}}}\right)^2}, \label{eqn:AnomCorrTrick}
\end{multline}
and identically for $\langle \hat{a}^{\dagger}(k,t_c)\hat{a}^{\dagger}(k',t_c)\hat{a}(k+2k_L,t_c)\hat{a}(k'-2k_L,t_c) \rangle$, where again $\Delta{k} \equiv k + k'$. For completeness, we note that the derivation of 
Equation (\ref{eqn:AnomCorrTrick}) can be performed by following similar logic and steps to the BB correlation (as per Appendix~\ref{app:ShortTime_g2}).

Substitution of Eqs.~(\ref{eqn:G2_BellReplace}) and (\ref{eqn:AnomCorrTrick}) into Eq.~(\ref{eqn:G2_kkdash_t4}) enables a straightforward evaluation of the integral in Eq.~(\ref{eqn:C_ij}) and 
thus the integrated pair-correlation function is found to be:
\begin{equation}
 \mathcal{C}_{12}(\phi_L,\phi_R) = \bar{n}^2\frac{\mathcal{K}^2}{4} + \frac{\bar{n}^2h_{\mathrm{BB}}}{2}\mathcal{B}_{\mathrm{BB}}\left[ 1 + \mathrm{cos}(\phi_L-\phi_R) \right] ,
\end{equation}
where $\mathcal{B}_{\mathrm{BB}}$ is defined as per Eq.~(\ref{eqn:B_BB}) in the main text.

It is then straightforward to follow the remainder of Ref.~\cite{RLS_Bell_2015} and arrive at the result:
\begin{equation}
 \mathcal{S} = 2\sqrt{2} \frac{h_{\mathrm{BB}} \mathcal{B}_{\mathrm{BB}}}{\frac{\mathcal{K}^2}{2} + h_{\mathrm{BB}} \mathcal{B}_{\mathrm{BB}}} .
\end{equation}
The final key factor differentiating this result to that of Ref.~\cite{RLS_Bell_2015} is the ignorance of any overall complex phase in the general form of Eq.~(\ref{eqn:AnomCorrTrick}). The 
presence of this phase factor is known to significantly degrade the predicted $\mathcal{S}$ due to an effective drift from the optimal set of phases $(\phi_L,\phi_R)$ if not correctly compensated for in the 
timing of the interferometric scheme. For the case of a BEC, this problem is known to increase with the size of the integration regions. Given that the typical size of integration region will likely be large 
(relative to the total size of the twin-beams in momentum space) to compensate for the broadening of the BB correlation function, this issue is expected to be important. However, improved characterisation of this
phase-factor will require more sophisticated (analytic) models of the collision process, similar to the perturbative technique \cite{Trippenbach_Perturbative_2008} applied in Ref.~\cite{RLS_Bell_2015}.

\section{positive-$P$ method \label{app:PosP_eqns}}
As a comparison to the short-time analytic model we simulate the dynamics of the system using the positive-$P$-representation \cite{gardiner2004quantum}. This effectively amounts to mapping the quantum field operators 
to corresponding complex stochastic fields, $\hat{\psi}_i(x,t) \rightarrow \psi_i(x,t)$ and $\hat{\psi}^{\dagger}_i(x,t) \rightarrow \tilde{\psi}_i(x,t)$, which evolve according to the Ito 
stochastic differential equations:
\begin{widetext}
\begin{eqnarray}
 \frac{d\psi_0}{dt} & = & \frac{i\hbar}{2m}\nabla^2\psi_0 - \frac{i}{\hbar} \left[ g_{00}\tilde{\psi}_0\psi_0 + 2g_{01}\tilde{\psi}_1\psi_1\right]\psi_0 - \frac{ig_{01}}{\hbar}\tilde{\psi}_0\psi^2_1
 + \sqrt{\frac{-i}{\hbar}(g_{00}\psi^2_0 + g_{01}\psi^2_1)}\xi_1 + \sqrt{\frac{-ig_{01}}{2\hbar}\psi_0\psi_1}\eta_1 , \notag \\
 \frac{d\psi_1}{dt} & = & \frac{i\hbar}{2m}\nabla^2\psi_1 - \frac{i}{\hbar} \left[ g_{11}\tilde{\psi}_1\psi_1 + 2g_{01}\tilde{\psi}_0\psi_0 + \hbar\omega_y \right]\psi_1 - \frac{ig_{01}}{\hbar}\tilde{\psi}_1\psi^2_0 + \sqrt{\frac{-i}{\hbar}(g_{11}\psi^2_1 + g_{01}\psi^2_0)}\xi_2 
 + \sqrt{\frac{-ig_{01}}{2\hbar}\psi_0\psi_1}\eta^*_1 , \notag \\
 \frac{d\tilde{\psi}_0}{dt} & = & -\frac{i\hbar}{2m}\nabla^2\tilde{\psi}_0 + \frac{i}{\hbar} \left[ g_{00}\psi_0\tilde{\psi}_0 + 2g_{01}\tilde{\psi}_1\psi_1 \right] \tilde{\psi}_0 + \frac{ig_{01}}{\hbar}\tilde{\psi}^2_1\psi_0 + \sqrt{\frac{i}{\hbar}(g_{00}\tilde{\psi}^2_0 + g_{01}\tilde{\psi}^2_1)}\xi_3 + \sqrt{\frac{ig_{01}}{2\hbar}\tilde{\psi}_0\tilde{\psi}_1}\eta_2 , \notag \\
 \frac{d\tilde{\psi}_1}{dt} & = & -\frac{i\hbar}{2m}\nabla^2\tilde{\psi}_1 + \frac{i}{\hbar} \left[ g_{11}\tilde{\psi}_1\psi_1 + 2g_{01}\tilde{\psi}_0\psi_0 + \hbar\omega_y \right]\tilde{\psi}_1 
 + \frac{ig_{01}}{\hbar}\tilde{\psi}^2_0\psi_1 + \sqrt{\frac{i}{\hbar}(g_{11}\tilde{\psi}^2_1 + g_{01}\tilde{\psi}^2_0)}\xi_4
 + \sqrt{\frac{ig_{01}}{2\hbar}\tilde{\psi}_0\tilde{\psi}_1}\eta^*_2 .
\end{eqnarray}
\end{widetext}
Here, $\xi_j(x,t)$ is a source of real Gaussian noise such that $\langle \xi_j(x,t) \rangle = 0$ and $\langle \xi_j(x,t) \xi_k(x',t') \rangle = \delta_{kj} \delta(x-x') \delta(t-t')$, 
while $\eta_j(x,t)$ is a source of complex Gaussian noise such that $\langle \eta_j(x,t) \rangle = 0$ and $\langle \eta^*_j(x,t) \eta_k(x',t') \rangle = \delta_{kj} \delta(x-x') \delta(t-t')$.

The initial condition for the excited quasi-condensate can be modelled within the positive-$P$ representation as $\psi_1(x,0) = \sqrt{\rho(x)}e^{i\varphi(x)}$ and 
$\tilde{\psi}_1(x,0) = \sqrt{\rho(x)}e^{-i\varphi(x)}$. Here, we have ignored density fluctuations such that $\rho(x)$ is the usual Thomas-Fermi density profile for a harmonic trap with $s$-wave interactions characterised by $g_{11}$, while the phase $\varphi(x)$ is sampled stochastically via
\begin{equation}
 \varphi(x) \equiv \sum_{j=1}^{\infty} \sqrt{\frac{(j+1/2)g_{11}}{2R_x\epsilon_j}} P_j\left(\frac{x}{R_x}\right) \left(\alpha_j + \beta_j \right) .
\end{equation}
This form is based off the form of the phase operator derived in Refs.~\cite{Petrov_1DqBEC_2000,Shevchenko_1992} where in the positive-$P$ representation the bosonic excitation operators are replaced by complex Gaussian random variables $\alpha_j$ and $\beta_j$ \cite{Olsen_PosPWig_2009} such that $\langle \alpha_j \rangle = 0$ ($\langle \beta_j \rangle = 0$) and $\langle \alpha^*_i\alpha_j \rangle = \delta_{ij} n_j$ ($\langle \beta^*_i \beta_j \rangle = \delta_{ij} n_j$).
The thermal mode occupation $n_j = 1/(e^{\epsilon_j/k_B T} - 1)$ is the usual Bose-Einstein distribution for phononic excitations at temperature $T$. For practical purposes, we truncate the sum 
for $n_j < 1$, implying that the positive-$P$ results will only be strictly valid at temperatures such that the low-energy phonon modes are highly occupied. 

Quantum mechanical expectation values are then obtained by appropriate averaging of the stochastic fields over a sufficiently large number of trajectories. For the positive-$P$ representation, averages over the stochastic fields correspond to normally-ordered expectation values of the field operators, such that $\langle (\hat{\psi}^{\dagger}_i)^m (\hat{\psi}_j)^n \rangle \equiv \langle (\tilde{\psi}_i)^m (\psi_j)^n \rangle_{\mathrm{stoch}}$.

%

\end{document}